\documentclass[%
 aip,
 amsmath,amssymb,
reprint,%
floatfix
]{revtex4-1}

\usepackage{graphicx}
\usepackage{dcolumn}
\usepackage{bm}

\usepackage[utf8]{inputenc}
\usepackage[T1]{fontenc}
\usepackage{mathptmx}
\usepackage[font=footnotesize]{caption}

\begin{document}

\preprint{}

\title[]{\large Strain effects on magnetic compensation and spin reorientation transition of Co/Gd synthetic ferrimagnets}

\author{Giovanni Masciocchi}
\email{gmascioc@uni-mainz.de}
 \affiliation{Institute of Physics, Johannes Gutenberg University Mainz, Staudingerweg 7, 55128 Mainz, Germany}
 \affiliation{Sensitec GmbH, Walter-Hallstein-Straße 24, 55130 Mainz, Germany}

\author{Thomas J. Kools}
\affiliation{Department of Applied Physics, Eindhoven University of Technology, Eindhoven, 5612 AZ, Netherlands}
 
\author{Pingzhi Li}%
\affiliation{Department of Applied Physics, Eindhoven University of Technology, Eindhoven, 5612 AZ, Netherlands}%

\author{ Adrien A. D. Petrillo}
\affiliation{Department of Applied Physics, Eindhoven University of Technology, Eindhoven, 5612 AZ, Netherlands}

\author{ Bert Koopmans}
\affiliation{Department of Applied Physics, Eindhoven University of Technology, Eindhoven, 5612 AZ, Netherlands}

\author{Reinoud Lavrijsen}
\affiliation{Department of Applied Physics, Eindhoven University of Technology, Eindhoven, 5612 AZ, Netherlands}

\author{ Andreas Kehlberger}
 \affiliation{Sensitec GmbH, Walter-Hallstein-Straße 24, 55130 Mainz, Germany}

\author{Mathias Kläui}
\affiliation{Institute of Physics, Johannes Gutenberg University Mainz, Staudingerweg 7, 55128 Mainz, Germany}
\date{\today}

\begin{abstract}

Synthetic ferrimagnets are an attractive materials class for spintronics as they provide access to all-optical switching of magnetization and, at the same time, allow for ultrafast domain wall motion at angular momentum compensation. In this work, we systematically study the effects of strain on the perpendicular magnetic anisotropy and magnetization compensation of Co/Gd and Co/Gd/Co/Gd synthetic ferrimagnets. Firstly, the spin reorientation transition of a bilayer system is investigated in wedge type samples, where we report an increase in the perpendicular magnetic anisotropy in the presence of in-plane strain. Using a model for magnetostatics and spin reorientation transition in this type of system, we confirm that the observed changes in anisotropy field are mainly due to the Co magnetoelastic anisotropy. Secondly, the magnetization compensation of a quadlayer is studied. We find that magnetization compensation of this synthetic ferrimagnetic system is not altered by external strain. This confirms the resilience of this material system against strain that may be induced during the integration process, making Co/Gd ferrimagnets suitable candidates for spintronics applications.



\end{abstract}

\maketitle



\section{Introduction}



Recent advances in spintronics have opened new possibilities for electronic applications beyond the CMOS standard. New concepts of high density and ultrafast non-volatile data storage have been proposed in magnetic memories\cite{endoh2020recent,parkin2008magnetic}. Throughout the years, magnetic memories have evolved \cite{tehrani2006status,garello2019spin} exploiting different geometries\cite{gu2022three} and new material platforms such as ferrimagnets\cite{yang2015domain} have been used to improve storage density\cite{tomasello2017performance}, reading and writing speed\cite{yang2019spin} and energy efficiency\cite{shao2022efficient,parkin2015memory}.  At the same time, single-pulse optical-switching (AOS) of magnetization has reduced the switching speed of the magnetization below ps timescale\cite{radu2011transient,ostler2012ultrafast,kimel2019writing,zhang2022all}. This bears promise for a new generation of ultrafast data buffering, in a single chip that integrates photonics with spintronics\cite{sobolewska2020integration, kim2022ferrimagnetic,aviles2019integration,lalieu2019integrating,becker2019out}. 

 Ferrimagnets are a class of magnets with unbalanced antiparallel-aligned sublattice moments. The compensation of the two inequivalent sublattices, combines the advantages of both antiferromagnets (antiparallel alignment of magnetic moments) and ferromagnets (finite Zeeman coupling and spin polarization)\cite{kim2022ferrimagnetic, kim2017fast}. Moreover, the drastic contrast between the two sublattices in non-adiabatic dynamics, could potentially accommodate AOS by a femtosecond laser pulse\cite{ostler2012ultrafast,kim2022ferrimagnetic}. Single-pulse AOS is typically  observed in rare earth–transition metal (RE–TM) ferrimagnetic alloys like  GdFeCo\cite{kim2017fast} or in multilayer synthetic ferrimagnet, such as Co/Gd and [Co/Tb]$_n$\cite{lalieu2017deterministic,aviles2020single}. 
 In particular, the one based on multilayer of Co/Gd is a good candidate for integrated opto-spintronics devices as it shows AOS - without the constrains on the composition as imposed by alloy system\cite{beens2019comparing,xu2017ultrafast} - and at the same time exhibits magnetic and angular momentum compensation, allowing  ultrafast domain wall motion \cite{pham2016very,li2022ultrafast}. For instance, the integration of Co/Gd synthetic ferrimagnets in an optically switchable magnetic tunnel junction has been recently reported\cite{wang2022picosecond}.


 When it comes to technological implementation, strain induced effects must be considered, which could be incurred from processing steps such as packaging and layer deposition\cite{windischmann1992intrinsic}. Intrinsic stresses and strain could affect the magnetic anisotropy  via changes to the spin-orbit coupling (SOC)\cite{twarowski1979magnetostriction} or to the magnetization compensation of ferrimagnets especially in RE–TM alloys\cite{chen2022modulating,zheng2022mechanically}. However, in spite of being omnipresent in applications\cite{tavassolizadeh2016tunnel,sahadevan2012biaxial,wang2018strain}, the effect of strain has not yet been explored in these materials. In this work, we present a systematic study of the effects of strain on Co/Gd synthetic ferrimagnets. By the application of external strain, using substrate bending, we investigate the impact of strain on the perpendicular magnetic anisotropy (PMA) and the magnetization compensation of [Co/Gd] and [Co/Gd]$_2$ multilayers, respectively. Using wedge samples in a bilayer system of Co/Gd and polar magneto-optic Kerr effect (pMOKE) measurements, we confirm that the PMA is increased by in-plane tensile strain and a negative magnetostriction is reported. By including the contribution of the strain-anisotropy for this system in a model for the magnetostatics, we show that the effects of strain on the magnetization are mainly due to the modification of the spin-orbit coupling within the magnetic layer and at the the Pt/Co interface that increases the magnetic anisotropy via magnetoelastic coupling. Additionally, we find that the magnetization compensation point is not affected significantly by strain, as the magnetoelastic coupling affects the anisotropy rather than the magnetization of the two sublattices.
Our study explores the mechanisms that underlie the influence of strain on the magnetic anisotropy of Co/Gd ferrimagnets and contributes to a better understanding of the magnetoelastic effects of ferrimagnetic multilayers. These results could be employed for the optimization and development of spintronics devices, as well as for potential applications in fields such as magnetic memory and sensing.

\section{Methods and sample fabrication}
The samples were grown on a 1.5 $\mu$m thick, thermally oxidized SiOx on top of a 625 $\mu$m thick Si substrate by DC magnetron sputtering in a chamber with a typical base pressure of $5\times 10^{-9}$ mBar. To obtain a variable thickness (wedge) along the sample surface, a shutter in the close proximity of the sample is gradually closed during deposition. This allows to study the compensation and spin reorientation transition (SRT) within a single sample. Two types of samples are realized. Firstly, a bilayer of Ta(4 nm)/Pt(4)/ Co(0-2)/Gd(t$_{Gd}$)/TaN(4) with a constant Gd layer on top of a Co wedge is considered to study the SRT. In addition, a quadlayer of Ta(4)/Pt(4)/Co(0.6)/Gd(0-2)/Co(0.6)/Gd(1.5)/TaN(4), this time with a Gd wedge, is grown  to study the magnetization compensation.  

The magnetic properties of these wedge samples were investigated by pMOKE, where we are only sensitive to the out-of-plane (OOP) component of the Co magnetization at a wavelength of 658 nm. According to Fig. \ref{img_1} (a), the surface of the sample is scanned along the y-direction using a focused laser spot with a spot-size of $\simeq$250 $\mu$m diameter. Accordingly, the local magnetic properties and hysteresis loops can be measured as a function of layer thickness, with a negligible thickness gradient $<0.025$ nm within the used laser spot. All the measurements are performed at room temperature. To apply in-plane tensile strain to our multilayer, the substrate  is mechanically bent using a three-point method\cite{masciocchi2021strain}. A square sample of 1 by 1 cm is vertically constrained on two sides and pushed uniformly from below by a cylinder that has off-centered rotation axis. The device generates a tensile strain in the plane of the sample when the cylinder is rotated. As previously reported, the tensile strain is uniaxial along $x$ and uniform in the measured area of the sample. 
The in-plane strain magnitude is $0.1\%$ and has been measured with a strain gauge (RS PRO). More details about the strain generating device can be found in section S2 of the supplementary information.

\section{Results and discussion}

\subsection{Spin reorientation transition in Co/Gd bilayer}

\begin{figure*}
	\includegraphics[width=13cm]{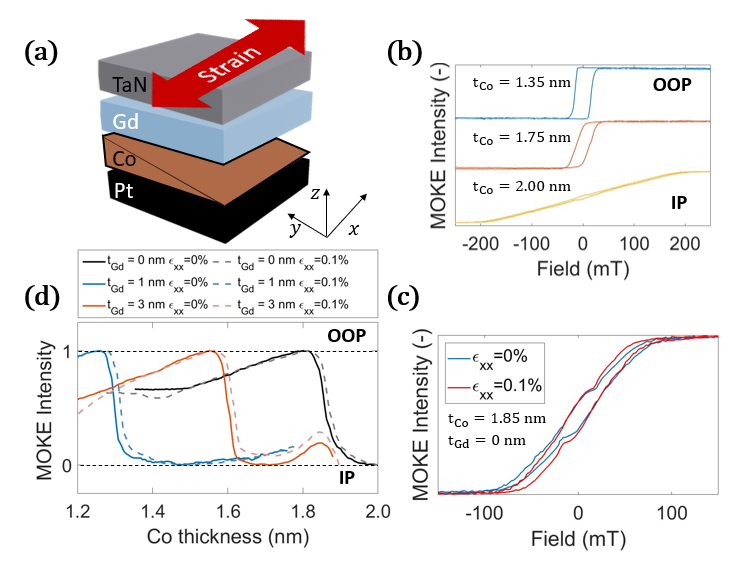}
	\caption{ (a) Sample sketch, red arrow indicates the direction of the applied strain. (b) Out of plane hysteresis loops of a Pt/Co/TaN stack for different Co thicknesses. (c) OOP hysteresis loops of Pt/Co(1.85 nm)/TaN  before (blue) and after (red) application of $0.1\%$ in-plane strain. (d) MOKE intensity scan at remanence (no applied field) of Pt/Co/Gd/TaN films along the Co wedge.  }
	\label{img_1}
\end{figure*}

The use of magnetic materials for high density data storage requires magnetic systems that are OOP magnetized\cite{chappert2007emergence,tudu2017recent}. In thin films, an OOP magnetic easy axis can be obtained  by magnetocrystalline anisotropy induced at the interface with heavy metal\cite{johnson1996magnetic,den1991magnetic}. In addition to that, strain  has been shown to affect the magnetic easy axis direction in systems with PMA\cite{kyuno1996theoretical}.  To understand the effect of external strain on Co/Gd systems with PMA, we investigate  bilayer samples consisting of Ta(4 nm)/Pt(4)/ Co(0-2)/Gd($t_{Gd}$)/TaN(4). Specifically, the Co thickness is varied between 0 and 2 nm over a few mm along the $y$ direction, whereas  $t_{Gd}$ is constant (as in Fig. \ref{img_1} (a)). In this system,  the balance between the interfacial anisotropy energy (magnetocrystalline anisotropy energy at the Pt/Co interface) and demagnetization energy determines the effective magnetic anisotropy. In such system, the demagnetization energy increases with the thickness of the Co magnetic layer, and consequently, the  magnetization will go from out-of-plane (OOP) to in-plane (IP). To probe the magnetization of our wedge sample, we record hysteresis loops from the pMOKE signal. We repeat the measurement moving the laser spot along the wedge in the y direction. Firstly, a sample where t$_{Gd}$=0 is considered. This measurement can be seen in Figs. \ref{img_1} (b) and (c). Fig. \ref{img_1} (b) reports the magnetic response of the Ta(4 nm)/Pt(4)/Co(0-2)/TaN(4) sample to an OOP magnetic film for different t$_{Co}$. The effective anisotropy $K_{eff}$ was estimated\cite{johnson1996magnetic} recording hysteresis loops with magnetic field applied OOP and IP and the corresponding anisotropy energy per unit area is $K_s=1.7$ mJ/m$^2$. For $t_{Co}=1.35$ nm the square-shaped loop indicates PMA with $K_{eff}$= 1.5(2)$\times 10^5$ J/m$^3$. A value of $M_{Co}=1.3$ MA/m was used in the calculation. As the thickness of Co is increased (moving the laser spot along the wedge direction - y) the remanence and squareness of the hysteresis loop decreases together with the PMA of the system. For $t_{Co}=2.00$ nm, the sample is IP magnetized and $K_{eff}=$ -0.8(2)$\times 10^5$ J/m$^3$ is negative. The OOP to IP transition occurs at $t_{Co}=1.85(2)$ nm in this system.  

To investigate the effects of externally applied in-plane strain, we repeat the measurement while the sample is mechanically bent.
The magnetization is coupled to the external strain and can be described by the expression for the anisotropy energy\cite{masciocchi2021strain}:

 \begin{equation} \label{strain_an}
K_{ME}=-\frac{3}{2}\lambda_s Y \epsilon,
\end{equation}

where $\lambda_s$ is the saturation magnetostriction, $Y$ is the Young's modulus and $\epsilon$ is the strain. If the strain in the film is non-zero, the magneto-elastic coupling of Co contributes in principle to the effective anisotropy. 
Accordingly, the total anisotropy $K_{eff}$ of the magnetic stack is expected to change in the presence of external strain. Fig. \ref{img_1} (c) shows the OOP hysteresis loops of Ta(4 nm)/Pt(4)/Co(1.85)/TaN(4) sample before (blue) and after (red) the application of $\epsilon_{xx}=0.1\%$. We observe that the anisotropy field is decreased after the application of in-plane strain. This happens because, in this system, the strain-induced magnetoelastic anisotropy $K_{ME}=0.02$ $mJ/m^2$ is positive, as we expect from a material with negative magnetostriction like Co\cite{hashimoto1989perpendicular,kyuno1996theoretical}. More details about the calculations of magnetoelastic anisotropy can be found in section S2 of the supplementary information. Accordingly, the PMA is increased by the applied strain, i.e. the system is expected to be OOP magnetized for thicker Co  if compared to samples without strain. 

After this preliminary study on Pt/Co systems, we focused our attention on the magnetostriction of Co/Gd multilayers. In Co-Gd alloys the magnetostriction has been reported to be strongly dependent on the composition\cite{twarowski1979magnetostriction,twarowski1981origin} due to the structural modification occurring with different atomic content. In contrast to this case, the effects of magnetostriction of a multilayer, are expected to be dependent on the magnetoelastic coupling of the individual layers\cite{masciocchi2022control}.

To study the magnetostriction of a Co/Gd multilayer, a constant layer of Gd on top of the Co wedge is added. To perform thickness dependent studies, a thickness $t_{Gd}=1$ nm and 3 nm is considered. In the bilayer system, the magnetization in the Gd layers is mainly induced at the interface with the Co layer, and couples anti-parallel the Co magnetization\cite{lalieu2017deterministic}. Accordingly,  $t_{Co}$ required to reach  SRT  is expected to change with increasing $t_{Gd}$\cite{kools2022magnetostatics}. To compare the SRT of Ta(4 nm)/Pt(4)/Co(0-2)/Gd($t_{Gd}$)/TaN(4) samples with different $t_{Gd}$  we performed remanent intensity scan along our Co wedge, in addition to hysteresis loop measurements. After the sample is saturated with an OOP magnetic field of 1T, we determine the thickness-dependent remanence from the pMOKE signal without external magnetic field. The remanent intensity scans are reported in Fig. \ref{img_1} (d).  As the pMOKE signal is mainly sensitive to the OOP component of Co magnetization, the normalized remanent intensity will drop to zero at the SRT, when the magnetization rotates IP. The SRT can be observed in Fig. \ref{img_1} (d) in samples with different thicknesses of Gd before and after the application of strain. As previously reported\cite{kools2022magnetostatics} the critical thickness $t_{Co}=t_c$ at which SRT occurs, changes significantly in the presence of a Gd layer. For all the considered samples, the in-plane strain shifts the OOP to IP transition towards larger Co thickness. This suggests that the effective magnetostriction of the  Co/Gd bilayer is negative and its value  $\lambda_s=-10(5)\times10^{-6}$ is not significantly altered by the presence of the Gd layer. 



\begin{figure}
	\includegraphics[width=8cm]{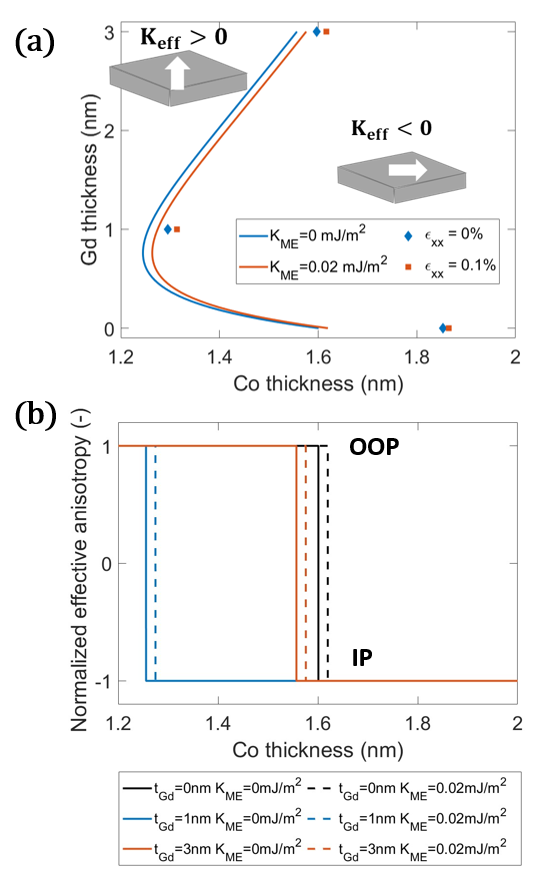}
	\caption{ (a) 2D phase diagram of the SRT of the a Ta(4nm)/Pt(4)/Co($t_{Co}$)/Gd(t$_{Gd}$)/TaN(4) stack as a function of $t_{Gd}$ (x) and $t_{Co}$ (y). The axes have been inverted for a better comparison with other figures. Blue diamonds and red  squares correspond to the experimental data, reported without and with strain applied, respectively. The solid lines indicate the calculated values using the model for the magnetostatics and Eq.  \ref{eq_str_eanis}. A magnetoelastic anisotropy $K_{ME}=0$ and 0.02 mJ/m$^2$ is considered, respectively, for the blue and orange curve. (b) Spin reorientation transition  of a Ta(4)/Pt(4)/Co($t_{Co}$)/Gd(t$_{Gd}$)/TaN(4) sample calculated for values of t$_{Gd}$=0, 1 and 3 nm and plotted as a function of $t_{Co}$. The SRT is represented here by a step function.  Solid and dashed lines consider $K_{ME}=0$ and 0.02 mJ/m$^2$, respectively. }
	\label{img_3}
\end{figure}

To obtain a quantitative understanding of the shape of the spin reorientation boundary, we employ an analytical model\cite{kools2022magnetostatics} describing the magnetostatic free energy of the anisotropy, which is zero at the SRT boundary. The first constituent energies of the model are the demagnetization energies of the Co layer

 \begin{equation} \label{demag_go}
E_{d,Co}=\frac{1}{2}\mu_0 \int_{0}^{y} M_{Co}^2 \,dq= \frac{1}{2}\mu_0 M_{Co}^2 y
\end{equation}

and of the Gd layer

 \begin{equation} \label{demag_co}
 \begin{split}
E_{d,Gd}=\frac{1}{2}\mu_0 \int_{0}^{x} M_{Gd}^2 exp(-2q/\lambda_{Gd}) \,dq=
\\
\frac{1}{4}\mu_0 M_{Gd}^2 \lambda_{Gd} \left(1-\exp\left(\frac{-2x}{\lambda_{Gd}}\right)\right)
\end{split}
\end{equation}

where $\lambda_{Gd}$ is the characteristic decay length of the Gd magnetization, which is induced at the Co/Gd interface, $M_{Co}$ is the magnetization of the Co  layer, $M_{Gd}$ is the effective Gd magnetization at the interface between Co and Gd and $x$ and $y$ are, respectively, the Gd and Co thicknesses in the diagram of Fig.\ref{img_3} (a). The plot axes in Fig.\ref{img_3} (a) have been inverted for a better comparison with the other figures.   The magnetocrystalline anisotropy is included with the term

 \begin{equation} \label{PMA_energy}
E_{K}=K_s - \Delta K \left(1-\exp\left(\frac{-2x}{\lambda_{K}}\right)\right),
\end{equation}

and it is also considered to decay with a characteristic decay length $\lambda_K$ and magnitude $\Delta K$. The second term in Eq. \ref{PMA_energy} phenomenologically addressed the experimentally observed decay in the effective anisotropy, which may be caused by sputter induced disordering of the Co\cite{bertero1993tem}. Using a numerical fit to the experimentally determined SRT, the parameters $\lambda_K$, $\lambda_{Gd}$ and $\Delta K$ for our Co/Gd bilayer are determined. All the other parameters were either experimentally measured or taken from literature and are reported in Table S.1 , section S1 of the supplementary information. In addition to the anisotropy term, and additional energy term $E_{mix}$ is included in the model. $E_{mix}$ takes into account the mixing at the magnetic layer interfaces where the local net magnetization is zero. More details about the expression for this term and the determination of the fitting parameters can be found in the supplementary information and in the work of Kools et al.\cite{kools2022magnetostatics}. In this model, the expression of the total free energy density per unit area is, considering all the terms mentioned so far:

 \begin{equation} \label{eq_strain_eanis}
E_{tot}= -E_K - E_{mix} +E_{d,Co}+E_{d,Gd}.
\end{equation}

 The magnetocrystalline anisotropy energy per unit area $K_s$, due to the Pt/Co interface is assumed constant.

Eq. \ref{eq_strain_eanis}, describing the total energy of a Ta(4nm)/Pt(4)/$ $Co($t_{Co}$)/Gd($t_{Gd}$)/TaN(4) sample, can be solved for y (t$_{Co}$) by imposing  $E_{tot}=0$ (spin reorientation transition). The solution for the SRT obtained with the model described above is reported in Fig. \ref{img_3} (a) with a blue solid line in a phase diagram where $t_{Gd}$ (x) and  $t_{Co}$ (y) are continuously varied from 0 to 3 nm and from 0 to 2 nm, respectively. Together with the calculations, the SRT  measured experimentally without externally applied strain is reported with blue diamonds in Fig. \ref{img_3} (a). The experimental data, follow well the  general trend of the calculations. Discrepancies between model and experimental values for $t_{Gd}=0$, might be due to additional mixing between the layers.

 To include the effects of strain, a magnetoelastic anisotropy $K_{ME}$ is added to Eq. \ref{eq_strain_eanis} that becomes

 \begin{equation} \label{eq_str_eanis}
E_{tot}= -E_K - E_{mix} - K_{ME}+E_{d,Co}+E_{d,Gd}.
\end{equation}

In our case $K_{ME}=0.02$ mJ/m$^2$  corresponds to the value of magnetoelastic anisotropy induced with $0.1\%$ externally applied in-plane strain in our experiments. As showed in Fig. \ref{img_1} (d), we do not observe significant changes to $K_{ME}$ with increasing $t_{Gd}$.
 Again considering the SRT-boundary  to be at $E_{tot} = 0$, the solution of Eq. \ref{eq_str_eanis} (that includes the magnetoelastic term)  is reported in Fig. \ref{img_3} (a) with an orange solid line. As expected from a material with negative magnetostriction, $K_{ME}$ sums to $K_s$ and the PMA is enhanced by in-plane strain. The SRT calculated including $K_{ME}$ to Eq.  \ref{eq_str_eanis} is consequently shifted to larger values of $t_{Co}$. This trend is in agreement with the experimentally determined SRT when and external strain $\epsilon_{xx}=0.1\%$ is applied (orange squares in Fig.\ref{img_3} (a)).

\begin{figure*}
	\includegraphics[width=16cm]{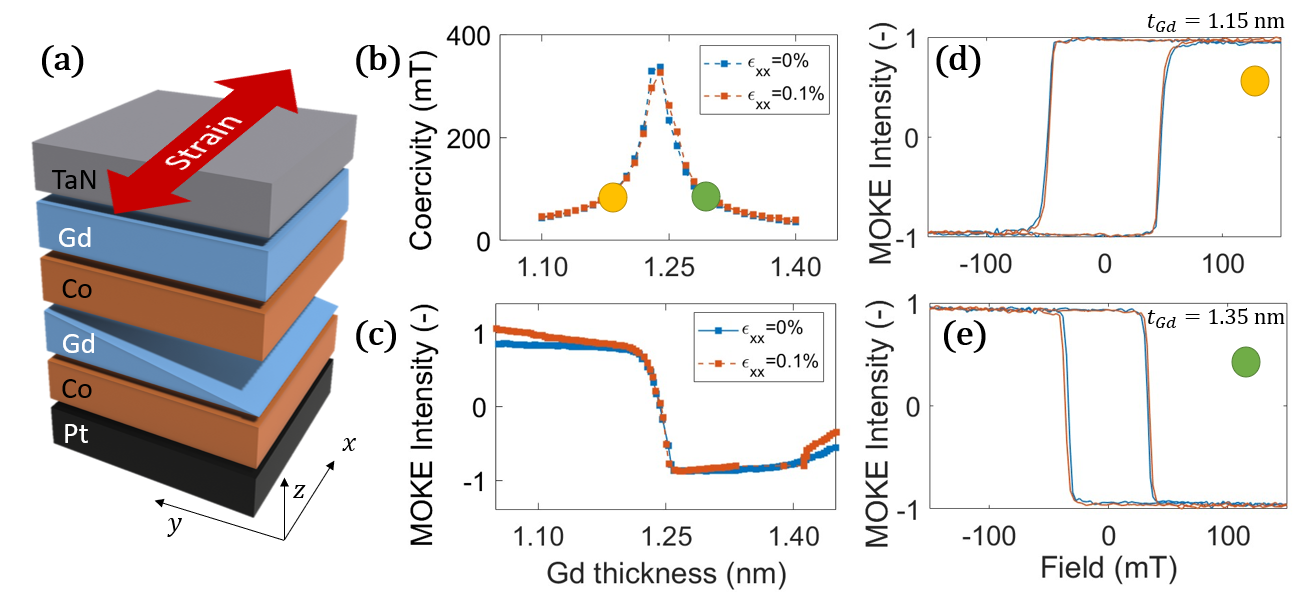}
	\caption{(a) Layerstack consisting of a Co/Gd quadlayer used to obtain magnetization compensation. (b) Coercivity and (c) remanent pMOKE intensity scan as a function of $t_{Gd}$. Measurements before (blue) and after (orange) application of in-plane strain are reported. (d) Hysteresis loops in the Co dominated and (e) Gd dominated state. Both curves with (orange) and without (blue) in-plane strain applied are shown. }
	\label{img_2}
\end{figure*}

Another way to visualize the SRT is solving Eq. \ref{eq_str_eanis}  for fixed values of $t_{Gd}$  and obtaining the critical thickness of $t_{Co}$ such that $E_{tot}=0$. Then, the SRT can be represented as a step function in the diagram of Fig. \ref{img_3} (b), analogue to the MOKE remanence scan shown in Fig. \ref{img_1} (d).  The  values of Gd thicknesses considered are $t_{Gd}=0$, 1 and 3 nm and are plotted in Fig. \ref{img_3} (b) with solid lines in black, blue and orange, in order. Solid lines consider $K_{ME}=0$ mJ/m$^2$. Dashed lines consider instead $K_{ME}=0.02$ mJ/m$^2$ in Fig. \ref{img_3} (b). The information contained here can be correlated to the experimental remanent intensity scan in Fig. \ref{img_1} (d).  Comparing Fig. \ref{img_3} (b) with Fig. \ref{img_1} (d), a similar behavior can be observed. Firstly we can note that the model predicts the SRT to shift when the thickness of the Gd layer is $t_{Gd}>0$. Secondly, we observe a  similar shift of the SRT point in Fig. \ref{img_3} (b) and  Fig. \ref{img_1} (d) due to the effect of magnetoelastic anisotropy and of the external strain, respectively. As we expect from a material with negative magnetostriction, $K_s$ adds to $K_{ME}$, therefore the PMA is increased and the Co/Gd bilayer stays OOP magnetized for thicker Co (corresponding to larger $E_{d,Co}$). We confirm that the major effect of strain on the Ta(4 nm)/Pt(4)/ Co(0-2)/Gd(t$_{Gd}$)/TaN(4) sample is the alteration of the PMA. Moreover, the estimated effective magnetostriction of the stack - $\lambda_s=-10(5)\times10^{-6}$ - is not significantly altered by the presence of the Gd layer in the thickness range considered. 

In this section, we examined the impact of in-plane strain on the effective PMA of a Co/Gd ferrimagnetic bilayer. Our results suggest negative magnetostriction of the stack for the investigated thickness values. We employ a recent model for the magnetostatics of these type of systems, where we include the effects of strain purely as magnetoelastic anisotropy. Our experimental findings are in good agreement with the predictions made by this model, providing deeper understanding of the response of this material platform to external strain.



\subsection{Magnetization compensation in quadlayer systems}

In ferrimagnets, magnetization compensation can be achieved. This occurs when  the net magnetization  $\vec{M_{tot}}=\vec{M_{Gd}}+\vec{M_{Co}}$ vanishes because the magnetization, coming from the two sub-lattices, is equal in magnitude and opposite in sign.

In recent studies, changes to the saturation magnetization in the presence of strain were reported in epitaxial films\cite{zheng2022mechanically} and rare earth free ferrimagnets\cite{chen2022modulating}. To study the effects of strain on magnetization compensation of synthetic ferrimagnets, we consider a quadlayer sample\cite{kools2022magnetostatics} consisting of Ta(4 nm)/Pt(4)/Co(0.6)/Gd(0-2)/Co(0.6)/Gd(1.5)/TaN(4) as schematically drawn in Fig. \ref{img_2} (a). In this case, the thickness of the bottom Gd layer is varied between 0 and 2 nm over a few mm, whereas all the other layers have constant thickness.  The reason for this choice  is that compared to the Co/Gd bilayer, the magnetic volume of the Co is doubled while the number of Co/Gd interfaces  where magnetization is induced in the Gd through direct exchange with the Co, is tripled. In this way magnetization compensation can be more readily achieved.

The growing thickness of Gd, increases the contribution of $\vec{M_{Gd}}$ to $\vec{M_{tot}}$. For this reason, some areas of the wedge sample will be Co-dominated (for $t_{Gd}<t_{comp}$) and other will be Gd-dominated (for $t_{Gd}>t_{comp}$) with $\vec{M_{tot}}=0$ at $t_{Gd}=t_{comp}$. Here, $t_{comp}$ is the thickness where magnetization compensation is obtained.
 At magnetization compensation two effects are expected: a divergence of the coercivity and a sign change in the remanent pMOKE signal (Kerr rotation, normalized to its value in absence of Gd). The measurements for coercivity and intensity are reported in Figs. \ref{img_2} (b) and (c), respectively. The coercivity data were extracted from hysteresis loops measured across the wedge direction (along y). The reason for the sign change in the pMOKE signal, is the alignment of the Gd magnetization along the field direction, in the Gd dominated regime. We report magnetization compensation in this quad-layer for $t_{Gd}=1.25$ nm.

In a similar fashion to what we have done investigating the PMA in the bilayer system, we repeat the experiment in the presence of $\epsilon_{xx}=$0.1$\%$ in-plane strain. The results are reported in orange in Fig. \ref{img_2} (b) and (c). Remarkably, the compensation point of the Co/Gd quadlayer is unchanged by the application of this externally applied strain.

Figs. \ref{img_2} (d) and (e) contain OOP hysteresis loops of  Ta(4 nm)/Pt(4)/Co(0.6)/Gd($t_{Gd}$)/Co(0.6)/Gd(1.5)/TaN(4) samples for $t_{Gd}=1.15$ nm and $t_{Gd}=1.35$ nm, respectively, and further show the effects of magnetization compensation. The sample is in this case OOP magnetized.  As the thickness of Gd is increased, the magnetization of the sample goes from Co  dominated (Fig. \ref{img_2} (d)) to Gd dominated (Fig. \ref{img_2} (e)). The inversion of hysteresis loops happens because for $t_{Gd}>1.25$ $nm$ the Co-magnetization aligns antiparallel to the field, leading to the change in sign of the pMOKE signal. When the measurement is repeated in the presence of $\epsilon_{xx}=0.1\%$ strain (orange line), no significant changes to the remanent intensity or coercivity are reported, if compared to the unstrained case (blue line). This suggests that magnetization compensation can be achieved in these multilayer systems in the presence of external strain and, most importantly, that the magnetization compensation point is unaffected.

To explain this, we can consider earlier studies about magnetostatics of these types of systems. As previously reported\cite{kools2022magnetostatics, pham2016very}, magnetization compensation is due to the balance in Co magnetization and the Gd magnetization, induced in the Gd at the Co/Gd interfaces. In-plane strain in multilayer samples with PMA modifies spin orbit coupling within one layer\cite{zhang1993magnetic}, thus altering the magnetocrystalline anisotropy energy of the system\cite{gopman2016strain}. On the other hand the total magnetic moment per unit area $\vec{M_{tot}}$ in synthetic ferrimagnets is obtained by integrating the
magnetization of the Co and Gd sublattices over the respective layer thicknesses. Accordingly, in a multilayer in-plane strain is not affecting the induced magnetic moment from the Co onto the Gd, thus not altering magnetization compensation. 


\section{Conclusions}

This work reveals the effect that external strain has on PMA and magnetization compensation of Co/Gd systems at room temperature. 
Growing wedge samples, where the thickness of one of the magnetic layers was varied, has allowed us to determine thickness dependent transition in the magnetostatics of this multilayer system. Deliberate in-plane strain was applied to the sample. In a bilayer Pt/Co/Gd system, we experimentally show that a sizable magnetoelastic coupling  changes the SRT in the presence of strain. The contribution of the strain-anisotropy for this system has been included in a model for the magnetostatics, describing the experimental observations well if an effective negative magnetostriction is considered. 
In a Pt/Co/Gd/Co/Gd quadlayer we obtain magnetization compensation of the two sub-lattices by varying the thickness of the bottom Gd layer. Here, we find that the application of in-plane strain does not affect the magnetization compensation. The induced magnetic moment from the Co onto the Gd, being an interface effect in a multilayer system, is not altered by such mechanical deformation.  
To conclude, this work provides a broad understanding of the magnetoelastic  properties of these multilayer systems. As PMA and magnetic compensation are maintained in the presence of externally applied  strain, this material system is a good candidate for technological implementation of ferrimagnets. 



\section*{Supplementary Material}
See supplementary material for magnetostatics model for the spin reorientation transition and for more details about the setup used for application of strain. 

\begin{acknowledgments}
 This project has received funding from the European Union’s Horizon 2020 research and innovation program  under  the  Marie  Skłodowska-Curie  grant  agreement  No  860060  “Magnetism  and  the effect of Electric Field” (MagnEFi), the Deutsche Forschungsgemeinschaft (DFG, German Research Foundation) - TRR 173 - 268565370 (project A01 and B02) and the Austrian Research Promotion Agency (FFG). The authors acknowledge support by the Max-Planck Graduate Centre with Johannes Gutenberg University.
\end{acknowledgments}

\section*{Author Declarations}
\subsection*{Conflict of interest }
The authors have no conflicts to disclose.

\section*{Data Sharing Policy }
The data that support the findings of this study are available from the corresponding author upon reasonable reques.





\nocite{*}
\bibliography{bibliography}

\begin{thebibliography}{47}%
\makeatletter
\providecommand \@ifxundefined [1]{%
 \@ifx{#1\undefined}
}%
\providecommand \@ifnum [1]{%
 \ifnum #1\expandafter \@firstoftwo
 \else \expandafter \@secondoftwo
 \fi
}%
\providecommand \@ifx [1]{%
 \ifx #1\expandafter \@firstoftwo
 \else \expandafter \@secondoftwo
 \fi
}%
\providecommand \natexlab [1]{#1}%
\providecommand \enquote  [1]{``#1''}%
\providecommand \bibnamefont  [1]{#1}%
\providecommand \bibfnamefont [1]{#1}%
\providecommand \citenamefont [1]{#1}%
\providecommand \href@noop [0]{\@secondoftwo}%
\providecommand \href [0]{\begingroup \@sanitize@url \@href}%
\providecommand \@href[1]{\@@startlink{#1}\@@href}%
\providecommand \@@href[1]{\endgroup#1\@@endlink}%
\providecommand \@sanitize@url [0]{\catcode `\\12\catcode `\$12\catcode
  `\&12\catcode `\#12\catcode `\^12\catcode `\_12\catcode `\%12\relax}%
\providecommand \@@startlink[1]{}%
\providecommand \@@endlink[0]{}%
\providecommand \url  [0]{\begingroup\@sanitize@url \@url }%
\providecommand \@url [1]{\endgroup\@href {#1}{\urlprefix }}%
\providecommand \urlprefix  [0]{URL }%
\providecommand \Eprint [0]{\href }%
\providecommand \doibase [0]{http://dx.doi.org/}%
\providecommand \selectlanguage [0]{\@gobble}%
\providecommand \bibinfo  [0]{\@secondoftwo}%
\providecommand \bibfield  [0]{\@secondoftwo}%
\providecommand \translation [1]{[#1]}%
\providecommand \BibitemOpen [0]{}%
\providecommand \bibitemStop [0]{}%
\providecommand \bibitemNoStop [0]{.\EOS\space}%
\providecommand \EOS [0]{\spacefactor3000\relax}%
\providecommand \BibitemShut  [1]{\csname bibitem#1\endcsname}%
\let\auto@bib@innerbib\@empty
\bibitem [{\citenamefont {Endoh}\ \emph {et~al.}(2020)\citenamefont {Endoh},
  \citenamefont {Honjo}, \citenamefont {Nishioka},\ and\ \citenamefont
  {Ikeda}}]{endoh2020recent}%
  \BibitemOpen
  \bibfield  {author} {\bibinfo {author} {\bibfnamefont {T.}~\bibnamefont
  {Endoh}}, \bibinfo {author} {\bibfnamefont {H.}~\bibnamefont {Honjo}},
  \bibinfo {author} {\bibfnamefont {K.}~\bibnamefont {Nishioka}}, \ and\
  \bibinfo {author} {\bibfnamefont {S.}~\bibnamefont {Ikeda}},\ }\bibfield
  {title} {\enquote {\bibinfo {title} {Recent progresses in {STT-MRAM and
  SOT-MRAM for next generation MRAM}},}\ }in\ \href@noop {} {\emph {\bibinfo
  {booktitle} {2020 IEEE Symposium on VLSI Technology}}}\ (\bibinfo
  {organization} {IEEE},\ \bibinfo {year} {2020})\ pp.\ \bibinfo {pages}
  {1--2}\BibitemShut {NoStop}%
\bibitem [{\citenamefont {Parkin}, \citenamefont {Hayashi},\ and\ \citenamefont
  {Thomas}(2008)}]{parkin2008magnetic}%
  \BibitemOpen
  \bibfield  {author} {\bibinfo {author} {\bibfnamefont {S.~S.}\ \bibnamefont
  {Parkin}}, \bibinfo {author} {\bibfnamefont {M.}~\bibnamefont {Hayashi}}, \
  and\ \bibinfo {author} {\bibfnamefont {L.}~\bibnamefont {Thomas}},\
  }\bibfield  {title} {\enquote {\bibinfo {title} {Magnetic domain-wall
  racetrack memory},}\ }\href@noop {} {\bibfield  {journal} {\bibinfo
  {journal} {Science}\ }\textbf {\bibinfo {volume} {320}},\ \bibinfo {pages}
  {190--194} (\bibinfo {year} {2008})}\BibitemShut {NoStop}%
\bibitem [{\citenamefont {Tehrani}(2006)}]{tehrani2006status}%
  \BibitemOpen
  \bibfield  {author} {\bibinfo {author} {\bibfnamefont {S.}~\bibnamefont
  {Tehrani}},\ }\bibfield  {title} {\enquote {\bibinfo {title} {Status and
  outlook of {MRAM} memory technology},}\ }in\ \href@noop {} {\emph {\bibinfo
  {booktitle} {2006 International Electron Devices Meeting}}}\ (\bibinfo
  {organization} {IEEE},\ \bibinfo {year} {2006})\ pp.\ \bibinfo {pages}
  {1--4}\BibitemShut {NoStop}%
\bibitem [{\citenamefont {Garello}, \citenamefont {Yasin},\ and\ \citenamefont
  {Kar}(2019)}]{garello2019spin}%
  \BibitemOpen
  \bibfield  {author} {\bibinfo {author} {\bibfnamefont {K.}~\bibnamefont
  {Garello}}, \bibinfo {author} {\bibfnamefont {F.}~\bibnamefont {Yasin}}, \
  and\ \bibinfo {author} {\bibfnamefont {G.~S.}\ \bibnamefont {Kar}},\
  }\bibfield  {title} {\enquote {\bibinfo {title} {Spin-orbit torque {MRAM} for
  ultrafast embedded memories: From fundamentals to large scale technology
  integration},}\ }in\ \href@noop {} {\emph {\bibinfo {booktitle} {2019 IEEE
  11th International Memory Workshop (IMW)}}}\ (\bibinfo {organization}
  {IEEE},\ \bibinfo {year} {2019})\ pp.\ \bibinfo {pages} {1--4}\BibitemShut
  {NoStop}%
\bibitem [{\citenamefont {Gu}\ \emph {et~al.}(2022)\citenamefont {Gu},
  \citenamefont {Guan}, \citenamefont {Hazra}, \citenamefont {Deniz},
  \citenamefont {Migliorini}, \citenamefont {Zhang},\ and\ \citenamefont
  {Parkin}}]{gu2022three}%
  \BibitemOpen
  \bibfield  {author} {\bibinfo {author} {\bibfnamefont {K.}~\bibnamefont
  {Gu}}, \bibinfo {author} {\bibfnamefont {Y.}~\bibnamefont {Guan}}, \bibinfo
  {author} {\bibfnamefont {B.~K.}\ \bibnamefont {Hazra}}, \bibinfo {author}
  {\bibfnamefont {H.}~\bibnamefont {Deniz}}, \bibinfo {author} {\bibfnamefont
  {A.}~\bibnamefont {Migliorini}}, \bibinfo {author} {\bibfnamefont
  {W.}~\bibnamefont {Zhang}}, \ and\ \bibinfo {author} {\bibfnamefont {S.~S.}\
  \bibnamefont {Parkin}},\ }\bibfield  {title} {\enquote {\bibinfo {title}
  {Three-dimensional racetrack memory devices designed from freestanding
  magnetic heterostructures},}\ }\href@noop {} {\bibfield  {journal} {\bibinfo
  {journal} {Nature Nanotechnology}\ }\textbf {\bibinfo {volume} {17}},\
  \bibinfo {pages} {1065--1071} (\bibinfo {year} {2022})}\BibitemShut {NoStop}%
\bibitem [{\citenamefont {Yang}, \citenamefont {Ryu},\ and\ \citenamefont
  {Parkin}(2015)}]{yang2015domain}%
  \BibitemOpen
  \bibfield  {author} {\bibinfo {author} {\bibfnamefont {S.-H.}\ \bibnamefont
  {Yang}}, \bibinfo {author} {\bibfnamefont {K.-S.}\ \bibnamefont {Ryu}}, \
  and\ \bibinfo {author} {\bibfnamefont {S.}~\bibnamefont {Parkin}},\
  }\bibfield  {title} {\enquote {\bibinfo {title} {{Domain-wall velocities of
  up to 750 m s- 1 driven by exchange-coupling torque in synthetic
  antiferromagnets}},}\ }\href@noop {} {\bibfield  {journal} {\bibinfo
  {journal} {Nature nanotechnology}\ }\textbf {\bibinfo {volume} {10}},\
  \bibinfo {pages} {221--226} (\bibinfo {year} {2015})}\BibitemShut {NoStop}%
\bibitem [{\citenamefont {Tomasello}\ \emph {et~al.}(2017)\citenamefont
  {Tomasello}, \citenamefont {Puliafito}, \citenamefont {Martinez},
  \citenamefont {Manchon}, \citenamefont {Ricci}, \citenamefont {Carpentieri},\
  and\ \citenamefont {Finocchio}}]{tomasello2017performance}%
  \BibitemOpen
  \bibfield  {author} {\bibinfo {author} {\bibfnamefont {R.}~\bibnamefont
  {Tomasello}}, \bibinfo {author} {\bibfnamefont {V.}~\bibnamefont
  {Puliafito}}, \bibinfo {author} {\bibfnamefont {E.}~\bibnamefont {Martinez}},
  \bibinfo {author} {\bibfnamefont {A.}~\bibnamefont {Manchon}}, \bibinfo
  {author} {\bibfnamefont {M.}~\bibnamefont {Ricci}}, \bibinfo {author}
  {\bibfnamefont {M.}~\bibnamefont {Carpentieri}}, \ and\ \bibinfo {author}
  {\bibfnamefont {G.}~\bibnamefont {Finocchio}},\ }\bibfield  {title} {\enquote
  {\bibinfo {title} {Performance of synthetic antiferromagnetic racetrack
  memory: domain wall versus skyrmion},}\ }\href@noop {} {\bibfield  {journal}
  {\bibinfo  {journal} {Journal of Physics D: Applied Physics}\ }\textbf
  {\bibinfo {volume} {50}},\ \bibinfo {pages} {325302} (\bibinfo {year}
  {2017})}\BibitemShut {NoStop}%
\bibitem [{\citenamefont {Yang}\ \emph {et~al.}(2019)\citenamefont {Yang},
  \citenamefont {Garg}, \citenamefont {Phung}, \citenamefont {Rettner},\ and\
  \citenamefont {Hughes}}]{yang2019spin}%
  \BibitemOpen
  \bibfield  {author} {\bibinfo {author} {\bibfnamefont {S.-H.}\ \bibnamefont
  {Yang}}, \bibinfo {author} {\bibfnamefont {C.}~\bibnamefont {Garg}}, \bibinfo
  {author} {\bibfnamefont {T.}~\bibnamefont {Phung}}, \bibinfo {author}
  {\bibfnamefont {C.}~\bibnamefont {Rettner}}, \ and\ \bibinfo {author}
  {\bibfnamefont {B.}~\bibnamefont {Hughes}},\ }\bibfield  {title} {\enquote
  {\bibinfo {title} {Spin-orbit torque driven one-bit magnetic racetrack
  devices-memory and neuromorphic applications},}\ }in\ \href@noop {} {\emph
  {\bibinfo {booktitle} {2019 International Symposium on VLSI Technology,
  Systems and Application (VLSI-TSA)}}}\ (\bibinfo {organization} {IEEE},\
  \bibinfo {year} {2019})\ pp.\ \bibinfo {pages} {1--2}\BibitemShut {NoStop}%
\bibitem [{\citenamefont {Shao}, \citenamefont {Wang},\ and\ \citenamefont
  {Yang}(2022)}]{shao2022efficient}%
  \BibitemOpen
  \bibfield  {author} {\bibinfo {author} {\bibfnamefont {Q.}~\bibnamefont
  {Shao}}, \bibinfo {author} {\bibfnamefont {Z.}~\bibnamefont {Wang}}, \ and\
  \bibinfo {author} {\bibfnamefont {J.~J.}\ \bibnamefont {Yang}},\ }\bibfield
  {title} {\enquote {\bibinfo {title} {Efficient {AI with MRAM}},}\ }\href@noop
  {} {\bibfield  {journal} {\bibinfo  {journal} {Nature Electronics}\ }\textbf
  {\bibinfo {volume} {5}},\ \bibinfo {pages} {67--68} (\bibinfo {year}
  {2022})}\BibitemShut {NoStop}%
\bibitem [{\citenamefont {Parkin}\ and\ \citenamefont
  {Yang}(2015)}]{parkin2015memory}%
  \BibitemOpen
  \bibfield  {author} {\bibinfo {author} {\bibfnamefont {S.}~\bibnamefont
  {Parkin}}\ and\ \bibinfo {author} {\bibfnamefont {S.-H.}\ \bibnamefont
  {Yang}},\ }\bibfield  {title} {\enquote {\bibinfo {title} {Memory on the
  racetrack},}\ }\href@noop {} {\bibfield  {journal} {\bibinfo  {journal}
  {Nature nanotechnology}\ }\textbf {\bibinfo {volume} {10}},\ \bibinfo {pages}
  {195--198} (\bibinfo {year} {2015})}\BibitemShut {NoStop}%
\bibitem [{\citenamefont {Radu}\ \emph {et~al.}(2011)\citenamefont {Radu},
  \citenamefont {Vahaplar}, \citenamefont {Stamm}, \citenamefont {Kachel},
  \citenamefont {Pontius}, \citenamefont {D{\"u}rr}, \citenamefont {Ostler},
  \citenamefont {Barker}, \citenamefont {Evans}, \citenamefont {Chantrell}
  \emph {et~al.}}]{radu2011transient}%
  \BibitemOpen
  \bibfield  {author} {\bibinfo {author} {\bibfnamefont {I.}~\bibnamefont
  {Radu}}, \bibinfo {author} {\bibfnamefont {K.}~\bibnamefont {Vahaplar}},
  \bibinfo {author} {\bibfnamefont {C.}~\bibnamefont {Stamm}}, \bibinfo
  {author} {\bibfnamefont {T.}~\bibnamefont {Kachel}}, \bibinfo {author}
  {\bibfnamefont {N.}~\bibnamefont {Pontius}}, \bibinfo {author} {\bibfnamefont
  {H.}~\bibnamefont {D{\"u}rr}}, \bibinfo {author} {\bibfnamefont
  {T.}~\bibnamefont {Ostler}}, \bibinfo {author} {\bibfnamefont
  {J.}~\bibnamefont {Barker}}, \bibinfo {author} {\bibfnamefont
  {R.}~\bibnamefont {Evans}}, \bibinfo {author} {\bibfnamefont
  {R.}~\bibnamefont {Chantrell}},  \emph {et~al.},\ }\bibfield  {title}
  {\enquote {\bibinfo {title} {Transient ferromagnetic-like state mediating
  ultrafast reversal of antiferromagnetically coupled spins},}\ }\href@noop {}
  {\bibfield  {journal} {\bibinfo  {journal} {Nature}\ }\textbf {\bibinfo
  {volume} {472}},\ \bibinfo {pages} {205--208} (\bibinfo {year}
  {2011})}\BibitemShut {NoStop}%
\bibitem [{\citenamefont {Ostler}\ \emph {et~al.}(2012)\citenamefont {Ostler},
  \citenamefont {Barker}, \citenamefont {Evans}, \citenamefont {Chantrell},
  \citenamefont {Atxitia}, \citenamefont {Chubykalo-Fesenko}, \citenamefont
  {El~Moussaoui}, \citenamefont {Le~Guyader}, \citenamefont {Mengotti},
  \citenamefont {Heyderman} \emph {et~al.}}]{ostler2012ultrafast}%
  \BibitemOpen
  \bibfield  {author} {\bibinfo {author} {\bibfnamefont {T.}~\bibnamefont
  {Ostler}}, \bibinfo {author} {\bibfnamefont {J.}~\bibnamefont {Barker}},
  \bibinfo {author} {\bibfnamefont {R.}~\bibnamefont {Evans}}, \bibinfo
  {author} {\bibfnamefont {R.}~\bibnamefont {Chantrell}}, \bibinfo {author}
  {\bibfnamefont {U.}~\bibnamefont {Atxitia}}, \bibinfo {author} {\bibfnamefont
  {O.}~\bibnamefont {Chubykalo-Fesenko}}, \bibinfo {author} {\bibfnamefont
  {S.}~\bibnamefont {El~Moussaoui}}, \bibinfo {author} {\bibfnamefont
  {L.}~\bibnamefont {Le~Guyader}}, \bibinfo {author} {\bibfnamefont
  {E.}~\bibnamefont {Mengotti}}, \bibinfo {author} {\bibfnamefont
  {L.}~\bibnamefont {Heyderman}},  \emph {et~al.},\ }\bibfield  {title}
  {\enquote {\bibinfo {title} {Ultrafast heating as a sufficient stimulus for
  magnetization reversal in a ferrimagnet},}\ }\href@noop {} {\bibfield
  {journal} {\bibinfo  {journal} {Nature communications}\ }\textbf {\bibinfo
  {volume} {3}},\ \bibinfo {pages} {1--6} (\bibinfo {year} {2012})}\BibitemShut
  {NoStop}%
\bibitem [{\citenamefont {Kimel}\ and\ \citenamefont
  {Li}(2019)}]{kimel2019writing}%
  \BibitemOpen
  \bibfield  {author} {\bibinfo {author} {\bibfnamefont {A.~V.}\ \bibnamefont
  {Kimel}}\ and\ \bibinfo {author} {\bibfnamefont {M.}~\bibnamefont {Li}},\
  }\bibfield  {title} {\enquote {\bibinfo {title} {Writing magnetic memory with
  ultrashort light pulses},}\ }\href@noop {} {\bibfield  {journal} {\bibinfo
  {journal} {Nature Reviews Materials}\ }\textbf {\bibinfo {volume} {4}},\
  \bibinfo {pages} {189--200} (\bibinfo {year} {2019})}\BibitemShut {NoStop}%
\bibitem [{\citenamefont {Zhang}\ \emph {et~al.}(2022)\citenamefont {Zhang},
  \citenamefont {Chung}, \citenamefont {Li}, \citenamefont {Wang},
  \citenamefont {Wang}, \citenamefont {Huey}, \citenamefont {Yang},
  \citenamefont {Goldberger}, \citenamefont {Yao},\ and\ \citenamefont
  {Zhang}}]{zhang2022all}%
  \BibitemOpen
  \bibfield  {author} {\bibinfo {author} {\bibfnamefont {P.}~\bibnamefont
  {Zhang}}, \bibinfo {author} {\bibfnamefont {T.-F.}\ \bibnamefont {Chung}},
  \bibinfo {author} {\bibfnamefont {Q.}~\bibnamefont {Li}}, \bibinfo {author}
  {\bibfnamefont {S.}~\bibnamefont {Wang}}, \bibinfo {author} {\bibfnamefont
  {Q.}~\bibnamefont {Wang}}, \bibinfo {author} {\bibfnamefont {W.~L.}\
  \bibnamefont {Huey}}, \bibinfo {author} {\bibfnamefont {S.}~\bibnamefont
  {Yang}}, \bibinfo {author} {\bibfnamefont {J.~E.}\ \bibnamefont
  {Goldberger}}, \bibinfo {author} {\bibfnamefont {J.}~\bibnamefont {Yao}}, \
  and\ \bibinfo {author} {\bibfnamefont {X.}~\bibnamefont {Zhang}},\ }\bibfield
   {title} {\enquote {\bibinfo {title} {All-optical switching of magnetization
  in atomically thin {CrI$_3$}},}\ }\href@noop {} {\bibfield  {journal}
  {\bibinfo  {journal} {Nature materials}\ }\textbf {\bibinfo {volume} {21}},\
  \bibinfo {pages} {1373--1378} (\bibinfo {year} {2022})}\BibitemShut {NoStop}%
\bibitem [{\citenamefont {Sobolewska}\ \emph {et~al.}(2020)\citenamefont
  {Sobolewska}, \citenamefont {Pelloux-Prayer}, \citenamefont {Becker},
  \citenamefont {Li}, \citenamefont {Davies}, \citenamefont {Kr{\"u}ckel},
  \citenamefont {F{\'e}lix}, \citenamefont {Olivier}, \citenamefont {Sousa},
  \citenamefont {Prejbeanu} \emph {et~al.}}]{sobolewska2020integration}%
  \BibitemOpen
  \bibfield  {author} {\bibinfo {author} {\bibfnamefont {E.~K.}\ \bibnamefont
  {Sobolewska}}, \bibinfo {author} {\bibfnamefont {J.}~\bibnamefont
  {Pelloux-Prayer}}, \bibinfo {author} {\bibfnamefont {H.}~\bibnamefont
  {Becker}}, \bibinfo {author} {\bibfnamefont {G.}~\bibnamefont {Li}}, \bibinfo
  {author} {\bibfnamefont {C.~S.}\ \bibnamefont {Davies}}, \bibinfo {author}
  {\bibfnamefont {C.}~\bibnamefont {Kr{\"u}ckel}}, \bibinfo {author}
  {\bibfnamefont {L.~A.}\ \bibnamefont {F{\'e}lix}}, \bibinfo {author}
  {\bibfnamefont {A.}~\bibnamefont {Olivier}}, \bibinfo {author} {\bibfnamefont
  {R.~C.}\ \bibnamefont {Sousa}}, \bibinfo {author} {\bibfnamefont {I.-L.}\
  \bibnamefont {Prejbeanu}},  \emph {et~al.},\ }\bibfield  {title} {\enquote
  {\bibinfo {title} {Integration platform for optical switching of magnetic
  elements},}\ }in\ \href@noop {} {\emph {\bibinfo {booktitle} {Active Photonic
  Platforms XII}}},\ Vol.\ \bibinfo {volume} {11461}\ (\bibinfo {organization}
  {SPIE},\ \bibinfo {year} {2020})\ pp.\ \bibinfo {pages} {54--72}\BibitemShut
  {NoStop}%
\bibitem [{\citenamefont {Kim}\ \emph {et~al.}(2022)\citenamefont {Kim},
  \citenamefont {Beach}, \citenamefont {Lee}, \citenamefont {Ono},
  \citenamefont {Rasing},\ and\ \citenamefont {Yang}}]{kim2022ferrimagnetic}%
  \BibitemOpen
  \bibfield  {author} {\bibinfo {author} {\bibfnamefont {S.~K.}\ \bibnamefont
  {Kim}}, \bibinfo {author} {\bibfnamefont {G.~S.}\ \bibnamefont {Beach}},
  \bibinfo {author} {\bibfnamefont {K.-J.}\ \bibnamefont {Lee}}, \bibinfo
  {author} {\bibfnamefont {T.}~\bibnamefont {Ono}}, \bibinfo {author}
  {\bibfnamefont {T.}~\bibnamefont {Rasing}}, \ and\ \bibinfo {author}
  {\bibfnamefont {H.}~\bibnamefont {Yang}},\ }\bibfield  {title} {\enquote
  {\bibinfo {title} {Ferrimagnetic spintronics},}\ }\href@noop {} {\bibfield
  {journal} {\bibinfo  {journal} {Nature Materials}\ }\textbf {\bibinfo
  {volume} {21}},\ \bibinfo {pages} {24--34} (\bibinfo {year}
  {2022})}\BibitemShut {NoStop}%
\bibitem [{\citenamefont {Avil{\'e}s-F{\'e}lix}\ \emph
  {et~al.}(2019)\citenamefont {Avil{\'e}s-F{\'e}lix}, \citenamefont
  {{\'A}lvaro-G{\'o}mez}, \citenamefont {Li}, \citenamefont {Davies},
  \citenamefont {Olivier}, \citenamefont {Rubio-Roy}, \citenamefont {Auffret},
  \citenamefont {Kirilyuk}, \citenamefont {Kimel}, \citenamefont {Rasing} \emph
  {et~al.}}]{aviles2019integration}%
  \BibitemOpen
  \bibfield  {author} {\bibinfo {author} {\bibfnamefont {L.}~\bibnamefont
  {Avil{\'e}s-F{\'e}lix}}, \bibinfo {author} {\bibfnamefont {L.}~\bibnamefont
  {{\'A}lvaro-G{\'o}mez}}, \bibinfo {author} {\bibfnamefont {G.}~\bibnamefont
  {Li}}, \bibinfo {author} {\bibfnamefont {C.}~\bibnamefont {Davies}}, \bibinfo
  {author} {\bibfnamefont {A.}~\bibnamefont {Olivier}}, \bibinfo {author}
  {\bibfnamefont {M.}~\bibnamefont {Rubio-Roy}}, \bibinfo {author}
  {\bibfnamefont {S.}~\bibnamefont {Auffret}}, \bibinfo {author} {\bibfnamefont
  {A.}~\bibnamefont {Kirilyuk}}, \bibinfo {author} {\bibfnamefont
  {A.}~\bibnamefont {Kimel}}, \bibinfo {author} {\bibfnamefont
  {T.}~\bibnamefont {Rasing}},  \emph {et~al.},\ }\bibfield  {title} {\enquote
  {\bibinfo {title} {Integration of {Tb/Co} multilayers within optically
  switchable perpendicular magnetic tunnel junctions},}\ }\href@noop {}
  {\bibfield  {journal} {\bibinfo  {journal} {Aip Advances}\ }\textbf {\bibinfo
  {volume} {9}},\ \bibinfo {pages} {125328} (\bibinfo {year}
  {2019})}\BibitemShut {NoStop}%
\bibitem [{\citenamefont {Lalieu}, \citenamefont {Lavrijsen},\ and\
  \citenamefont {Koopmans}(2019)}]{lalieu2019integrating}%
  \BibitemOpen
  \bibfield  {author} {\bibinfo {author} {\bibfnamefont {M.~L.}\ \bibnamefont
  {Lalieu}}, \bibinfo {author} {\bibfnamefont {R.}~\bibnamefont {Lavrijsen}}, \
  and\ \bibinfo {author} {\bibfnamefont {B.}~\bibnamefont {Koopmans}},\
  }\bibfield  {title} {\enquote {\bibinfo {title} {Integrating all-optical
  switching with spintronics},}\ }\href@noop {} {\bibfield  {journal} {\bibinfo
   {journal} {Nature communications}\ }\textbf {\bibinfo {volume} {10}},\
  \bibinfo {pages} {110} (\bibinfo {year} {2019})}\BibitemShut {NoStop}%
\bibitem [{\citenamefont {Becker}\ \emph {et~al.}(2019)\citenamefont {Becker},
  \citenamefont {Kr{\"u}ckel}, \citenamefont {Van~Thourhout},\ and\
  \citenamefont {Heck}}]{becker2019out}%
  \BibitemOpen
  \bibfield  {author} {\bibinfo {author} {\bibfnamefont {H.}~\bibnamefont
  {Becker}}, \bibinfo {author} {\bibfnamefont {C.~J.}\ \bibnamefont
  {Kr{\"u}ckel}}, \bibinfo {author} {\bibfnamefont {D.}~\bibnamefont
  {Van~Thourhout}}, \ and\ \bibinfo {author} {\bibfnamefont {M.~J.}\
  \bibnamefont {Heck}},\ }\bibfield  {title} {\enquote {\bibinfo {title}
  {Out-of-plane focusing grating couplers for silicon photonics integration
  with optical {MRAM} technology},}\ }\href@noop {} {\bibfield  {journal}
  {\bibinfo  {journal} {IEEE Journal of Selected Topics in Quantum
  Electronics}\ }\textbf {\bibinfo {volume} {26}},\ \bibinfo {pages} {1--8}
  (\bibinfo {year} {2019})}\BibitemShut {NoStop}%
\bibitem [{\citenamefont {Kim}\ \emph {et~al.}(2017)\citenamefont {Kim},
  \citenamefont {Kim}, \citenamefont {Hirata}, \citenamefont {Oh},
  \citenamefont {Tono}, \citenamefont {Kim}, \citenamefont {Okuno},
  \citenamefont {Ham}, \citenamefont {Kim}, \citenamefont {Go} \emph
  {et~al.}}]{kim2017fast}%
  \BibitemOpen
  \bibfield  {author} {\bibinfo {author} {\bibfnamefont {K.-J.}\ \bibnamefont
  {Kim}}, \bibinfo {author} {\bibfnamefont {S.~K.}\ \bibnamefont {Kim}},
  \bibinfo {author} {\bibfnamefont {Y.}~\bibnamefont {Hirata}}, \bibinfo
  {author} {\bibfnamefont {S.-H.}\ \bibnamefont {Oh}}, \bibinfo {author}
  {\bibfnamefont {T.}~\bibnamefont {Tono}}, \bibinfo {author} {\bibfnamefont
  {D.-H.}\ \bibnamefont {Kim}}, \bibinfo {author} {\bibfnamefont
  {T.}~\bibnamefont {Okuno}}, \bibinfo {author} {\bibfnamefont {W.~S.}\
  \bibnamefont {Ham}}, \bibinfo {author} {\bibfnamefont {S.}~\bibnamefont
  {Kim}}, \bibinfo {author} {\bibfnamefont {G.}~\bibnamefont {Go}},  \emph
  {et~al.},\ }\bibfield  {title} {\enquote {\bibinfo {title} {Fast domain wall
  motion in the vicinity of the angular momentum compensation temperature of
  ferrimagnets},}\ }\href@noop {} {\bibfield  {journal} {\bibinfo  {journal}
  {Nature materials}\ }\textbf {\bibinfo {volume} {16}},\ \bibinfo {pages}
  {1187--1192} (\bibinfo {year} {2017})}\BibitemShut {NoStop}%
\bibitem [{\citenamefont {Lalieu}\ \emph {et~al.}(2017)\citenamefont {Lalieu},
  \citenamefont {Peeters}, \citenamefont {Haenen}, \citenamefont {Lavrijsen},\
  and\ \citenamefont {Koopmans}}]{lalieu2017deterministic}%
  \BibitemOpen
  \bibfield  {author} {\bibinfo {author} {\bibfnamefont {M.}~\bibnamefont
  {Lalieu}}, \bibinfo {author} {\bibfnamefont {M.}~\bibnamefont {Peeters}},
  \bibinfo {author} {\bibfnamefont {S.}~\bibnamefont {Haenen}}, \bibinfo
  {author} {\bibfnamefont {R.}~\bibnamefont {Lavrijsen}}, \ and\ \bibinfo
  {author} {\bibfnamefont {B.}~\bibnamefont {Koopmans}},\ }\bibfield  {title}
  {\enquote {\bibinfo {title} {Deterministic all-optical switching of synthetic
  ferrimagnets using single femtosecond laser pulses},}\ }\href@noop {}
  {\bibfield  {journal} {\bibinfo  {journal} {Physical review B}\ }\textbf
  {\bibinfo {volume} {96}},\ \bibinfo {pages} {220411} (\bibinfo {year}
  {2017})}\BibitemShut {NoStop}%
\bibitem [{\citenamefont {Avil{\'e}s-F{\'e}lix}\ \emph
  {et~al.}(2020)\citenamefont {Avil{\'e}s-F{\'e}lix}, \citenamefont {Olivier},
  \citenamefont {Li}, \citenamefont {Davies}, \citenamefont
  {{\'A}lvaro-G{\'o}mez}, \citenamefont {Rubio-Roy}, \citenamefont {Auffret},
  \citenamefont {Kirilyuk}, \citenamefont {Kimel}, \citenamefont {Rasing} \emph
  {et~al.}}]{aviles2020single}%
  \BibitemOpen
  \bibfield  {author} {\bibinfo {author} {\bibfnamefont {L.}~\bibnamefont
  {Avil{\'e}s-F{\'e}lix}}, \bibinfo {author} {\bibfnamefont {A.}~\bibnamefont
  {Olivier}}, \bibinfo {author} {\bibfnamefont {G.}~\bibnamefont {Li}},
  \bibinfo {author} {\bibfnamefont {C.~S.}\ \bibnamefont {Davies}}, \bibinfo
  {author} {\bibfnamefont {L.}~\bibnamefont {{\'A}lvaro-G{\'o}mez}}, \bibinfo
  {author} {\bibfnamefont {M.}~\bibnamefont {Rubio-Roy}}, \bibinfo {author}
  {\bibfnamefont {S.}~\bibnamefont {Auffret}}, \bibinfo {author} {\bibfnamefont
  {A.}~\bibnamefont {Kirilyuk}}, \bibinfo {author} {\bibfnamefont
  {A.}~\bibnamefont {Kimel}}, \bibinfo {author} {\bibfnamefont
  {T.}~\bibnamefont {Rasing}},  \emph {et~al.},\ }\bibfield  {title} {\enquote
  {\bibinfo {title} {Single-shot all-optical switching of magnetization in
  {Tb/Co} multilayer-based electrodes},}\ }\href@noop {} {\bibfield  {journal}
  {\bibinfo  {journal} {Scientific reports}\ }\textbf {\bibinfo {volume}
  {10}},\ \bibinfo {pages} {1--8} (\bibinfo {year} {2020})}\BibitemShut
  {NoStop}%
\bibitem [{\citenamefont {Beens}\ \emph {et~al.}(2019)\citenamefont {Beens},
  \citenamefont {Lalieu}, \citenamefont {Deenen}, \citenamefont {Duine},\ and\
  \citenamefont {Koopmans}}]{beens2019comparing}%
  \BibitemOpen
  \bibfield  {author} {\bibinfo {author} {\bibfnamefont {M.}~\bibnamefont
  {Beens}}, \bibinfo {author} {\bibfnamefont {M.~L.}\ \bibnamefont {Lalieu}},
  \bibinfo {author} {\bibfnamefont {A.~J.}\ \bibnamefont {Deenen}}, \bibinfo
  {author} {\bibfnamefont {R.~A.}\ \bibnamefont {Duine}}, \ and\ \bibinfo
  {author} {\bibfnamefont {B.}~\bibnamefont {Koopmans}},\ }\bibfield  {title}
  {\enquote {\bibinfo {title} {Comparing all-optical switching in
  synthetic-ferrimagnetic multilayers and alloys},}\ }\href@noop {} {\bibfield
  {journal} {\bibinfo  {journal} {Physical Review B}\ }\textbf {\bibinfo
  {volume} {100}},\ \bibinfo {pages} {220409} (\bibinfo {year}
  {2019})}\BibitemShut {NoStop}%
\bibitem [{\citenamefont {Xu}\ \emph {et~al.}(2017)\citenamefont {Xu},
  \citenamefont {Deb}, \citenamefont {Malinowski}, \citenamefont {Hehn},
  \citenamefont {Zhao},\ and\ \citenamefont {Mangin}}]{xu2017ultrafast}%
  \BibitemOpen
  \bibfield  {author} {\bibinfo {author} {\bibfnamefont {Y.}~\bibnamefont
  {Xu}}, \bibinfo {author} {\bibfnamefont {M.}~\bibnamefont {Deb}}, \bibinfo
  {author} {\bibfnamefont {G.}~\bibnamefont {Malinowski}}, \bibinfo {author}
  {\bibfnamefont {M.}~\bibnamefont {Hehn}}, \bibinfo {author} {\bibfnamefont
  {W.}~\bibnamefont {Zhao}}, \ and\ \bibinfo {author} {\bibfnamefont
  {S.}~\bibnamefont {Mangin}},\ }\bibfield  {title} {\enquote {\bibinfo {title}
  {Ultrafast magnetization manipulation using single femtosecond light and
  hot-electron pulses},}\ }\href@noop {} {\bibfield  {journal} {\bibinfo
  {journal} {Advanced Materials}\ }\textbf {\bibinfo {volume} {29}},\ \bibinfo
  {pages} {1703474} (\bibinfo {year} {2017})}\BibitemShut {NoStop}%
\bibitem [{\citenamefont {Pham}\ \emph {et~al.}(2016)\citenamefont {Pham},
  \citenamefont {Vogel}, \citenamefont {Sampaio}, \citenamefont
  {Va{\v{n}}atka}, \citenamefont {Rojas-S{\'a}nchez}, \citenamefont {Bonfim},
  \citenamefont {Chaves}, \citenamefont {Choueikani}, \citenamefont {Ohresser},
  \citenamefont {Otero} \emph {et~al.}}]{pham2016very}%
  \BibitemOpen
  \bibfield  {author} {\bibinfo {author} {\bibfnamefont {T.~H.}\ \bibnamefont
  {Pham}}, \bibinfo {author} {\bibfnamefont {J.}~\bibnamefont {Vogel}},
  \bibinfo {author} {\bibfnamefont {J.}~\bibnamefont {Sampaio}}, \bibinfo
  {author} {\bibfnamefont {M.}~\bibnamefont {Va{\v{n}}atka}}, \bibinfo {author}
  {\bibfnamefont {J.-C.}\ \bibnamefont {Rojas-S{\'a}nchez}}, \bibinfo {author}
  {\bibfnamefont {M.}~\bibnamefont {Bonfim}}, \bibinfo {author} {\bibfnamefont
  {D.}~\bibnamefont {Chaves}}, \bibinfo {author} {\bibfnamefont
  {F.}~\bibnamefont {Choueikani}}, \bibinfo {author} {\bibfnamefont
  {P.}~\bibnamefont {Ohresser}}, \bibinfo {author} {\bibfnamefont
  {E.}~\bibnamefont {Otero}},  \emph {et~al.},\ }\bibfield  {title} {\enquote
  {\bibinfo {title} {Very large domain wall velocities in{ Pt/Co/GdOx and
  Pt/Co/Gd trilayers with Dzyaloshinskii-Moriya} interaction},}\ }\href@noop {}
  {\bibfield  {journal} {\bibinfo  {journal} {EPL (Europhysics Letters)}\
  }\textbf {\bibinfo {volume} {113}},\ \bibinfo {pages} {67001} (\bibinfo
  {year} {2016})}\BibitemShut {NoStop}%
\bibitem [{\citenamefont {Li}\ \emph {et~al.}(2023)\citenamefont {Li},
  \citenamefont {Kools}, \citenamefont {Koopmans},\ and\ \citenamefont
  {Lavrijsen}}]{li2022ultrafast}%
  \BibitemOpen
  \bibfield  {author} {\bibinfo {author} {\bibfnamefont {P.}~\bibnamefont
  {Li}}, \bibinfo {author} {\bibfnamefont {T.~J.}\ \bibnamefont {Kools}},
  \bibinfo {author} {\bibfnamefont {B.}~\bibnamefont {Koopmans}}, \ and\
  \bibinfo {author} {\bibfnamefont {R.}~\bibnamefont {Lavrijsen}},\ }\bibfield
  {title} {\enquote {\bibinfo {title} {Ultrafast racetrack based on compensated
  {Co/Gd}-based synthetic ferrimagnet with {All-Optical Switching}},}\
  }\href@noop {} {\bibfield  {journal} {\bibinfo  {journal} {Advanced
  Electronic Materials}\ }\textbf {\bibinfo {volume} {9}},\ \bibinfo {pages}
  {2200613} (\bibinfo {year} {2023})}\BibitemShut {NoStop}%
\bibitem [{\citenamefont {Wang}\ \emph {et~al.}(2022)\citenamefont {Wang},
  \citenamefont {Cheng}, \citenamefont {Li}, \citenamefont {van Hees},
  \citenamefont {Liu}, \citenamefont {Cao}, \citenamefont {Lavrijsen},
  \citenamefont {Lin}, \citenamefont {Koopmans},\ and\ \citenamefont
  {Zhao}}]{wang2022picosecond}%
  \BibitemOpen
  \bibfield  {author} {\bibinfo {author} {\bibfnamefont {L.}~\bibnamefont
  {Wang}}, \bibinfo {author} {\bibfnamefont {H.}~\bibnamefont {Cheng}},
  \bibinfo {author} {\bibfnamefont {P.}~\bibnamefont {Li}}, \bibinfo {author}
  {\bibfnamefont {Y.~L.}\ \bibnamefont {van Hees}}, \bibinfo {author}
  {\bibfnamefont {Y.}~\bibnamefont {Liu}}, \bibinfo {author} {\bibfnamefont
  {K.}~\bibnamefont {Cao}}, \bibinfo {author} {\bibfnamefont {R.}~\bibnamefont
  {Lavrijsen}}, \bibinfo {author} {\bibfnamefont {X.}~\bibnamefont {Lin}},
  \bibinfo {author} {\bibfnamefont {B.}~\bibnamefont {Koopmans}}, \ and\
  \bibinfo {author} {\bibfnamefont {W.}~\bibnamefont {Zhao}},\ }\bibfield
  {title} {\enquote {\bibinfo {title} {Picosecond optospintronic tunnel
  junctions},}\ }\href@noop {} {\bibfield  {journal} {\bibinfo  {journal}
  {Proceedings of the National Academy of Sciences}\ }\textbf {\bibinfo
  {volume} {119}},\ \bibinfo {pages} {e2204732119} (\bibinfo {year}
  {2022})}\BibitemShut {NoStop}%
\bibitem [{\citenamefont {Windischmann}(1992)}]{windischmann1992intrinsic}%
  \BibitemOpen
  \bibfield  {author} {\bibinfo {author} {\bibfnamefont {H.}~\bibnamefont
  {Windischmann}},\ }\bibfield  {title} {\enquote {\bibinfo {title} {Intrinsic
  stress in sputter-deposited thin films},}\ }\href@noop {} {\bibfield
  {journal} {\bibinfo  {journal} {Critical Reviews in Solid State and Material
  Sciences}\ }\textbf {\bibinfo {volume} {17}},\ \bibinfo {pages} {547--596}
  (\bibinfo {year} {1992})}\BibitemShut {NoStop}%
\bibitem [{\citenamefont {Twarowski}\ and\ \citenamefont
  {Lachowicz}(1979)}]{twarowski1979magnetostriction}%
  \BibitemOpen
  \bibfield  {author} {\bibinfo {author} {\bibfnamefont {K.}~\bibnamefont
  {Twarowski}}\ and\ \bibinfo {author} {\bibfnamefont {H.}~\bibnamefont
  {Lachowicz}},\ }\bibfield  {title} {\enquote {\bibinfo {title}
  {Magnetostriction and anisotropy of amorphous {Gd-Co RF sputtered thin
  films}},}\ }\href@noop {} {\bibfield  {journal} {\bibinfo  {journal} {Journal
  of Applied Physics}\ }\textbf {\bibinfo {volume} {50}},\ \bibinfo {pages}
  {7722--7724} (\bibinfo {year} {1979})}\BibitemShut {NoStop}%
\bibitem [{\citenamefont {Chen}\ \emph {et~al.}(2022)\citenamefont {Chen},
  \citenamefont {Shi}, \citenamefont {Liu}, \citenamefont {Chen}, \citenamefont
  {Zhang},\ and\ \citenamefont {Mi}}]{chen2022modulating}%
  \BibitemOpen
  \bibfield  {author} {\bibinfo {author} {\bibfnamefont {Z.}~\bibnamefont
  {Chen}}, \bibinfo {author} {\bibfnamefont {X.}~\bibnamefont {Shi}}, \bibinfo
  {author} {\bibfnamefont {X.}~\bibnamefont {Liu}}, \bibinfo {author}
  {\bibfnamefont {X.}~\bibnamefont {Chen}}, \bibinfo {author} {\bibfnamefont
  {Z.}~\bibnamefont {Zhang}}, \ and\ \bibinfo {author} {\bibfnamefont
  {W.}~\bibnamefont {Mi}},\ }\bibfield  {title} {\enquote {\bibinfo {title}
  {Modulating saturation magnetization and topological{ Hall resistivity of
  flexible ferrimagnetic Mn4N films by bending strains}},}\ }\href@noop {}
  {\bibfield  {journal} {\bibinfo  {journal} {Journal of Applied Physics}\
  }\textbf {\bibinfo {volume} {132}},\ \bibinfo {pages} {233906} (\bibinfo
  {year} {2022})}\BibitemShut {NoStop}%
\bibitem [{\citenamefont {Zheng}, \citenamefont {Guan},\ and\ \citenamefont
  {Fan}(2022)}]{zheng2022mechanically}%
  \BibitemOpen
  \bibfield  {author} {\bibinfo {author} {\bibfnamefont {M.}~\bibnamefont
  {Zheng}}, \bibinfo {author} {\bibfnamefont {P.}~\bibnamefont {Guan}}, \ and\
  \bibinfo {author} {\bibfnamefont {H.}~\bibnamefont {Fan}},\ }\bibfield
  {title} {\enquote {\bibinfo {title} {Mechanically enhanced magnetism in
  {flexible semitransparent CuFe2O4/mica epitaxial }heterostructures},}\
  }\href@noop {} {\bibfield  {journal} {\bibinfo  {journal} {Applied Surface
  Science}\ }\textbf {\bibinfo {volume} {584}},\ \bibinfo {pages} {152586}
  (\bibinfo {year} {2022})}\BibitemShut {NoStop}%
\bibitem [{\citenamefont {Tavassolizadeh}\ \emph {et~al.}(2016)\citenamefont
  {Tavassolizadeh}, \citenamefont {Rott}, \citenamefont {Meier}, \citenamefont
  {Quandt}, \citenamefont {H{\"o}lscher}, \citenamefont {Reiss},\ and\
  \citenamefont {Meyners}}]{tavassolizadeh2016tunnel}%
  \BibitemOpen
  \bibfield  {author} {\bibinfo {author} {\bibfnamefont {A.}~\bibnamefont
  {Tavassolizadeh}}, \bibinfo {author} {\bibfnamefont {K.}~\bibnamefont
  {Rott}}, \bibinfo {author} {\bibfnamefont {T.}~\bibnamefont {Meier}},
  \bibinfo {author} {\bibfnamefont {E.}~\bibnamefont {Quandt}}, \bibinfo
  {author} {\bibfnamefont {H.}~\bibnamefont {H{\"o}lscher}}, \bibinfo {author}
  {\bibfnamefont {G.}~\bibnamefont {Reiss}}, \ and\ \bibinfo {author}
  {\bibfnamefont {D.}~\bibnamefont {Meyners}},\ }\bibfield  {title} {\enquote
  {\bibinfo {title} {Tunnel magnetoresistance sensors with magnetostrictive
  electrodes: Strain sensors},}\ }\href@noop {} {\bibfield  {journal} {\bibinfo
   {journal} {Sensors}\ }\textbf {\bibinfo {volume} {16}},\ \bibinfo {pages}
  {1902} (\bibinfo {year} {2016})}\BibitemShut {NoStop}%
\bibitem [{\citenamefont {Sahadevan}\ \emph {et~al.}(2012)\citenamefont
  {Sahadevan}, \citenamefont {Tiwari}, \citenamefont {Kalon}, \citenamefont
  {Bhatia}, \citenamefont {Saeys},\ and\ \citenamefont
  {Yang}}]{sahadevan2012biaxial}%
  \BibitemOpen
  \bibfield  {author} {\bibinfo {author} {\bibfnamefont {A.~M.}\ \bibnamefont
  {Sahadevan}}, \bibinfo {author} {\bibfnamefont {R.~K.}\ \bibnamefont
  {Tiwari}}, \bibinfo {author} {\bibfnamefont {G.}~\bibnamefont {Kalon}},
  \bibinfo {author} {\bibfnamefont {C.~S.}\ \bibnamefont {Bhatia}}, \bibinfo
  {author} {\bibfnamefont {M.}~\bibnamefont {Saeys}}, \ and\ \bibinfo {author}
  {\bibfnamefont {H.}~\bibnamefont {Yang}},\ }\bibfield  {title} {\enquote
  {\bibinfo {title} {Biaxial strain effect of spin dependent tunneling in{ MgO}
  magnetic tunnel junctions},}\ }\href@noop {} {\bibfield  {journal} {\bibinfo
  {journal} {Applied Physics Letters}\ }\textbf {\bibinfo {volume} {101}},\
  \bibinfo {pages} {042407} (\bibinfo {year} {2012})}\BibitemShut {NoStop}%
\bibitem [{\citenamefont {Wang}\ \emph {et~al.}(2018)\citenamefont {Wang},
  \citenamefont {Domann}, \citenamefont {Yu}, \citenamefont {Barra},
  \citenamefont {Wang},\ and\ \citenamefont {Carman}}]{wang2018strain}%
  \BibitemOpen
  \bibfield  {author} {\bibinfo {author} {\bibfnamefont {Q.}~\bibnamefont
  {Wang}}, \bibinfo {author} {\bibfnamefont {J.}~\bibnamefont {Domann}},
  \bibinfo {author} {\bibfnamefont {G.}~\bibnamefont {Yu}}, \bibinfo {author}
  {\bibfnamefont {A.}~\bibnamefont {Barra}}, \bibinfo {author} {\bibfnamefont
  {K.~L.}\ \bibnamefont {Wang}}, \ and\ \bibinfo {author} {\bibfnamefont
  {G.~P.}\ \bibnamefont {Carman}},\ }\bibfield  {title} {\enquote {\bibinfo
  {title} {Strain-mediated spin-orbit-torque switching for magnetic memory},}\
  }\href@noop {} {\bibfield  {journal} {\bibinfo  {journal} {Physical Review
  Applied}\ }\textbf {\bibinfo {volume} {10}},\ \bibinfo {pages} {034052}
  (\bibinfo {year} {2018})}\BibitemShut {NoStop}%
\bibitem [{\citenamefont {Masciocchi}\ \emph {et~al.}(2021)\citenamefont
  {Masciocchi}, \citenamefont {Fattouhi}, \citenamefont {Kehlberger},
  \citenamefont {Lopez-Diaz}, \citenamefont {Syskaki},\ and\ \citenamefont
  {Kl{\"a}ui}}]{masciocchi2021strain}%
  \BibitemOpen
  \bibfield  {author} {\bibinfo {author} {\bibfnamefont {G.}~\bibnamefont
  {Masciocchi}}, \bibinfo {author} {\bibfnamefont {M.}~\bibnamefont
  {Fattouhi}}, \bibinfo {author} {\bibfnamefont {A.}~\bibnamefont
  {Kehlberger}}, \bibinfo {author} {\bibfnamefont {L.}~\bibnamefont
  {Lopez-Diaz}}, \bibinfo {author} {\bibfnamefont {M.-A.}\ \bibnamefont
  {Syskaki}}, \ and\ \bibinfo {author} {\bibfnamefont {M.}~\bibnamefont
  {Kl{\"a}ui}},\ }\bibfield  {title} {\enquote {\bibinfo {title}
  {Strain-controlled domain wall injection into nanowires for sensor
  applications},}\ }\href@noop {} {\bibfield  {journal} {\bibinfo  {journal}
  {Journal of Applied Physics}\ }\textbf {\bibinfo {volume} {130}},\ \bibinfo
  {pages} {183903} (\bibinfo {year} {2021})}\BibitemShut {NoStop}%
\bibitem [{\citenamefont {Chappert}, \citenamefont {Fert},\ and\ \citenamefont
  {Van~Dau}(2007)}]{chappert2007emergence}%
  \BibitemOpen
  \bibfield  {author} {\bibinfo {author} {\bibfnamefont {C.}~\bibnamefont
  {Chappert}}, \bibinfo {author} {\bibfnamefont {A.}~\bibnamefont {Fert}}, \
  and\ \bibinfo {author} {\bibfnamefont {F.~N.}\ \bibnamefont {Van~Dau}},\
  }\bibfield  {title} {\enquote {\bibinfo {title} {The emergence of spin
  electronics in data storage},}\ }\href@noop {} {\bibfield  {journal}
  {\bibinfo  {journal} {Nature materials}\ }\textbf {\bibinfo {volume} {6}},\
  \bibinfo {pages} {813--823} (\bibinfo {year} {2007})}\BibitemShut {NoStop}%
\bibitem [{\citenamefont {Tudu}\ and\ \citenamefont
  {Tiwari}(2017)}]{tudu2017recent}%
  \BibitemOpen
  \bibfield  {author} {\bibinfo {author} {\bibfnamefont {B.}~\bibnamefont
  {Tudu}}\ and\ \bibinfo {author} {\bibfnamefont {A.}~\bibnamefont {Tiwari}},\
  }\bibfield  {title} {\enquote {\bibinfo {title} {Recent developments in
  perpendicular magnetic anisotropy thin films for data storage
  applications},}\ }\href@noop {} {\bibfield  {journal} {\bibinfo  {journal}
  {Vacuum}\ }\textbf {\bibinfo {volume} {146}},\ \bibinfo {pages} {329--341}
  (\bibinfo {year} {2017})}\BibitemShut {NoStop}%
\bibitem [{\citenamefont {Johnson}\ \emph {et~al.}(1996)\citenamefont
  {Johnson}, \citenamefont {Bloemen}, \citenamefont {Den~Broeder},\ and\
  \citenamefont {De~Vries}}]{johnson1996magnetic}%
  \BibitemOpen
  \bibfield  {author} {\bibinfo {author} {\bibfnamefont {M.}~\bibnamefont
  {Johnson}}, \bibinfo {author} {\bibfnamefont {P.}~\bibnamefont {Bloemen}},
  \bibinfo {author} {\bibfnamefont {F.}~\bibnamefont {Den~Broeder}}, \ and\
  \bibinfo {author} {\bibfnamefont {J.}~\bibnamefont {De~Vries}},\ }\bibfield
  {title} {\enquote {\bibinfo {title} {Magnetic anisotropy in metallic
  multilayers},}\ }\href@noop {} {\bibfield  {journal} {\bibinfo  {journal}
  {Reports on Progress in Physics}\ }\textbf {\bibinfo {volume} {59}},\
  \bibinfo {pages} {1409} (\bibinfo {year} {1996})}\BibitemShut {NoStop}%
\bibitem [{\citenamefont {Den~Broeder}, \citenamefont {Hoving},\ and\
  \citenamefont {Bloemen}(1991)}]{den1991magnetic}%
  \BibitemOpen
  \bibfield  {author} {\bibinfo {author} {\bibfnamefont {F.}~\bibnamefont
  {Den~Broeder}}, \bibinfo {author} {\bibfnamefont {W.}~\bibnamefont {Hoving}},
  \ and\ \bibinfo {author} {\bibfnamefont {P.}~\bibnamefont {Bloemen}},\
  }\bibfield  {title} {\enquote {\bibinfo {title} {Magnetic anisotropy of
  multilayers},}\ }\href@noop {} {\bibfield  {journal} {\bibinfo  {journal}
  {Journal of magnetism and magnetic materials}\ }\textbf {\bibinfo {volume}
  {93}},\ \bibinfo {pages} {562--570} (\bibinfo {year} {1991})}\BibitemShut
  {NoStop}%
\bibitem [{\citenamefont {Kyuno}\ \emph {et~al.}(1996)\citenamefont {Kyuno},
  \citenamefont {Ha}, \citenamefont {Yamamoto},\ and\ \citenamefont
  {Asano}}]{kyuno1996theoretical}%
  \BibitemOpen
  \bibfield  {author} {\bibinfo {author} {\bibfnamefont {K.}~\bibnamefont
  {Kyuno}}, \bibinfo {author} {\bibfnamefont {J.-G.}\ \bibnamefont {Ha}},
  \bibinfo {author} {\bibfnamefont {R.}~\bibnamefont {Yamamoto}}, \ and\
  \bibinfo {author} {\bibfnamefont {S.}~\bibnamefont {Asano}},\ }\bibfield
  {title} {\enquote {\bibinfo {title} {Theoretical study on the strain
  dependence of the magnetic anisotropy of {X/Co (X= Pt, Cu, Ag, and Au)}
  metallic multilayers},}\ }\href@noop {} {\bibfield  {journal} {\bibinfo
  {journal} {Journal of applied physics}\ }\textbf {\bibinfo {volume} {79}},\
  \bibinfo {pages} {7084--7089} (\bibinfo {year} {1996})}\BibitemShut {NoStop}%
\bibitem [{\citenamefont {Hashimoto}, \citenamefont {Ochiai},\ and\
  \citenamefont {Aso}(1989)}]{hashimoto1989perpendicular}%
  \BibitemOpen
  \bibfield  {author} {\bibinfo {author} {\bibfnamefont {S.}~\bibnamefont
  {Hashimoto}}, \bibinfo {author} {\bibfnamefont {Y.}~\bibnamefont {Ochiai}}, \
  and\ \bibinfo {author} {\bibfnamefont {K.}~\bibnamefont {Aso}},\ }\bibfield
  {title} {\enquote {\bibinfo {title} {Perpendicular magnetic anisotropy and
  magnetostriction of sputtered {Co/Pd and Co/Pt} multilayered films},}\
  }\href@noop {} {\bibfield  {journal} {\bibinfo  {journal} {Journal of applied
  physics}\ }\textbf {\bibinfo {volume} {66}},\ \bibinfo {pages} {4909--4916}
  (\bibinfo {year} {1989})}\BibitemShut {NoStop}%
\bibitem [{\citenamefont {Twarowski}\ \emph {et~al.}(1981)\citenamefont
  {Twarowski}, \citenamefont {Lachowicz}, \citenamefont {Gutowski},\ and\
  \citenamefont {Szymczak}}]{twarowski1981origin}%
  \BibitemOpen
  \bibfield  {author} {\bibinfo {author} {\bibfnamefont {K.}~\bibnamefont
  {Twarowski}}, \bibinfo {author} {\bibfnamefont {H.}~\bibnamefont
  {Lachowicz}}, \bibinfo {author} {\bibfnamefont {M.}~\bibnamefont {Gutowski}},
  \ and\ \bibinfo {author} {\bibfnamefont {H.}~\bibnamefont {Szymczak}},\
  }\bibfield  {title} {\enquote {\bibinfo {title} {On the origin of the
  perpendicular anisotropy and magnetostriction in amorphous {RF} sputtered {Gd
  Co} films},}\ }\href@noop {} {\bibfield  {journal} {\bibinfo  {journal}
  {physica status solidi (a)}\ }\textbf {\bibinfo {volume} {63}},\ \bibinfo
  {pages} {103--108} (\bibinfo {year} {1981})}\BibitemShut {NoStop}%
\bibitem [{\citenamefont {Masciocchi}\ \emph {et~al.}(2022)\citenamefont
  {Masciocchi}, \citenamefont {van~der Jagt}, \citenamefont {Syskaki},
  \citenamefont {Lamperti}, \citenamefont {Wolff}, \citenamefont {Lotnyk},
  \citenamefont {Langer}, \citenamefont {Kienle}, \citenamefont {Jakob},
  \citenamefont {Borie} \emph {et~al.}}]{masciocchi2022control}%
  \BibitemOpen
  \bibfield  {author} {\bibinfo {author} {\bibfnamefont {G.}~\bibnamefont
  {Masciocchi}}, \bibinfo {author} {\bibfnamefont {J.~W.}\ \bibnamefont
  {van~der Jagt}}, \bibinfo {author} {\bibfnamefont {M.-A.}\ \bibnamefont
  {Syskaki}}, \bibinfo {author} {\bibfnamefont {A.}~\bibnamefont {Lamperti}},
  \bibinfo {author} {\bibfnamefont {N.}~\bibnamefont {Wolff}}, \bibinfo
  {author} {\bibfnamefont {A.}~\bibnamefont {Lotnyk}}, \bibinfo {author}
  {\bibfnamefont {J.}~\bibnamefont {Langer}}, \bibinfo {author} {\bibfnamefont
  {L.}~\bibnamefont {Kienle}}, \bibinfo {author} {\bibfnamefont
  {G.}~\bibnamefont {Jakob}}, \bibinfo {author} {\bibfnamefont
  {B.}~\bibnamefont {Borie}},  \emph {et~al.},\ }\bibfield  {title} {\enquote
  {\bibinfo {title} {Control of magnetoelastic coupling in {Ni/Fe multilayers
  using He+ ion irradiation}},}\ }\href@noop {} {\bibfield  {journal} {\bibinfo
   {journal} {Applied Physics Letters}\ }\textbf {\bibinfo {volume} {121}},\
  \bibinfo {pages} {182401} (\bibinfo {year} {2022})}\BibitemShut {NoStop}%
\bibitem [{\citenamefont {Kools}\ \emph {et~al.}(2022)\citenamefont {Kools},
  \citenamefont {van Gurp}, \citenamefont {Koopmans},\ and\ \citenamefont
  {Lavrijsen}}]{kools2022magnetostatics}%
  \BibitemOpen
  \bibfield  {author} {\bibinfo {author} {\bibfnamefont {T.~J.}\ \bibnamefont
  {Kools}}, \bibinfo {author} {\bibfnamefont {M.~C.}\ \bibnamefont {van Gurp}},
  \bibinfo {author} {\bibfnamefont {B.}~\bibnamefont {Koopmans}}, \ and\
  \bibinfo {author} {\bibfnamefont {R.}~\bibnamefont {Lavrijsen}},\ }\bibfield
  {title} {\enquote {\bibinfo {title} {Magnetostatics of room temperature
  compensated {Co/Gd/Co/Gd}-based synthetic ferrimagnets},}\ }\href@noop {}
  {\bibfield  {journal} {\bibinfo  {journal} {Applied Physics Letters}\
  }\textbf {\bibinfo {volume} {121}},\ \bibinfo {pages} {242405} (\bibinfo
  {year} {2022})}\BibitemShut {NoStop}%
\bibitem [{\citenamefont {Bertero}\ \emph {et~al.}(1993)\citenamefont
  {Bertero}, \citenamefont {Hufnagel}, \citenamefont {Clemens},\ and\
  \citenamefont {Sinclair}}]{bertero1993tem}%
  \BibitemOpen
  \bibfield  {author} {\bibinfo {author} {\bibfnamefont {G.}~\bibnamefont
  {Bertero}}, \bibinfo {author} {\bibfnamefont {T.}~\bibnamefont {Hufnagel}},
  \bibinfo {author} {\bibfnamefont {B.}~\bibnamefont {Clemens}}, \ and\
  \bibinfo {author} {\bibfnamefont {R.}~\bibnamefont {Sinclair}},\ }\bibfield
  {title} {\enquote {\bibinfo {title} {{TEM analysis of Co-Gd and Co-Gd
  multilayer structures}},}\ }\href@noop {} {\bibfield  {journal} {\bibinfo
  {journal} {Journal of materials research}\ }\textbf {\bibinfo {volume} {8}},\
  \bibinfo {pages} {771--774} (\bibinfo {year} {1993})}\BibitemShut {NoStop}%
\bibitem [{\citenamefont {Zhang}\ \emph {et~al.}(1993)\citenamefont {Zhang},
  \citenamefont {Krishnan}, \citenamefont {Lee},\ and\ \citenamefont
  {Farrow}}]{zhang1993magnetic}%
  \BibitemOpen
  \bibfield  {author} {\bibinfo {author} {\bibfnamefont {B.}~\bibnamefont
  {Zhang}}, \bibinfo {author} {\bibfnamefont {K.~M.}\ \bibnamefont {Krishnan}},
  \bibinfo {author} {\bibfnamefont {C.}~\bibnamefont {Lee}}, \ and\ \bibinfo
  {author} {\bibfnamefont {R.}~\bibnamefont {Farrow}},\ }\bibfield  {title}
  {\enquote {\bibinfo {title} {Magnetic anisotropy and lattice strain in
  {Co/Pt} multilayers},}\ }\href@noop {} {\bibfield  {journal} {\bibinfo
  {journal} {Journal of applied physics}\ }\textbf {\bibinfo {volume} {73}},\
  \bibinfo {pages} {6198--6200} (\bibinfo {year} {1993})}\BibitemShut {NoStop}%
\bibitem [{\citenamefont {Gopman}\ \emph {et~al.}(2016)\citenamefont {Gopman},
  \citenamefont {Dennis}, \citenamefont {Chen}, \citenamefont {Iunin},
  \citenamefont {Finkel}, \citenamefont {Staruch},\ and\ \citenamefont
  {Shull}}]{gopman2016strain}%
  \BibitemOpen
  \bibfield  {author} {\bibinfo {author} {\bibfnamefont {D.~B.}\ \bibnamefont
  {Gopman}}, \bibinfo {author} {\bibfnamefont {C.~L.}\ \bibnamefont {Dennis}},
  \bibinfo {author} {\bibfnamefont {P.}~\bibnamefont {Chen}}, \bibinfo {author}
  {\bibfnamefont {Y.~L.}\ \bibnamefont {Iunin}}, \bibinfo {author}
  {\bibfnamefont {P.}~\bibnamefont {Finkel}}, \bibinfo {author} {\bibfnamefont
  {M.}~\bibnamefont {Staruch}}, \ and\ \bibinfo {author} {\bibfnamefont
  {R.~D.}\ \bibnamefont {Shull}},\ }\bibfield  {title} {\enquote {\bibinfo
  {title} {Strain-assisted magnetization reversal in {Co/Ni} multilayers with
  perpendicular magnetic anisotropy},}\ }\href@noop {} {\bibfield  {journal}
  {\bibinfo  {journal} {Scientific reports}\ }\textbf {\bibinfo {volume} {6}},\
  \bibinfo {pages} {1--8} (\bibinfo {year} {2016})}\BibitemShut {NoStop}%
\end{thebibliography}%

\newpage



\end{document}


\preprint{}

\title[\textbf{Suppl. material} - Strain effects on magnetic compensation and spin reorientation transition of ...]{Supplementary material -  Strain effects on magnetic compensation and spin reorientation transition of Co/Gd synthetic ferrimagnets}

\date{\today}

\maketitle

\subsection*{\textbf{S1} - Magnetostatics model for the Spin Reorientation Transition}

In the expression of the free energy, a term describing the effect of intermixing at the multilayer system interfaces is added. The expression is 

 \begin{equation} \tag{S.1}\label{demag_mix}
 \begin{array}{l}
E_{mix}=\frac{1}{2}\mu_0 \int_{0}^{a_0x} M_{Co}^2+ \left( M_{Gd}exp(-q/\lambda_{Gd}) \right)^2 \,dq= \\ \frac{1}{2}\mu_0 a_0 M_{Co}^2 x  + \frac{1}{4}\mu_0 \lambda_{Gd} M_{Gd}^2 \left(1-\exp\left(\frac{-2a_0x}{\lambda_{Gd}}\right)\right).
\end{array}
\end{equation}

  Therefore, the total energy $E_{tot}= -E_K - E_{mix} - K_{ME}+E_{d,Co}+E_{d,Gd}$, including all the terms will be:

 \begin{equation} \tag{S.2} \label{total_mix}
 \begin{array}{l}
E_{tot}=-K_s + \Delta K \left(1-\exp\left(\frac{-2x}{\lambda_{K}}\right)\right) -K_{ME} - \frac{1}{2}\mu_0 a_0 M_{Co}^2 x   \\ -\frac{1}{4}\mu_0 \lambda_{Gd} M_{Gd}^2 \left(1-\exp\left(\frac{-2a_0x}{\lambda_{Gd}}\right)\right)+ \frac{1}{2}\mu_0 M_{Co}^2 y  \\ + \frac{1}{4}\mu_0 M_{Gd}^2 \lambda_{Gd} \left(1-\exp\left(\frac{-2x}{\lambda_{Gd}}\right)\right).
\end{array}
\end{equation}

\begin{figure}[h!]
\centering\includegraphics[width=12cm]{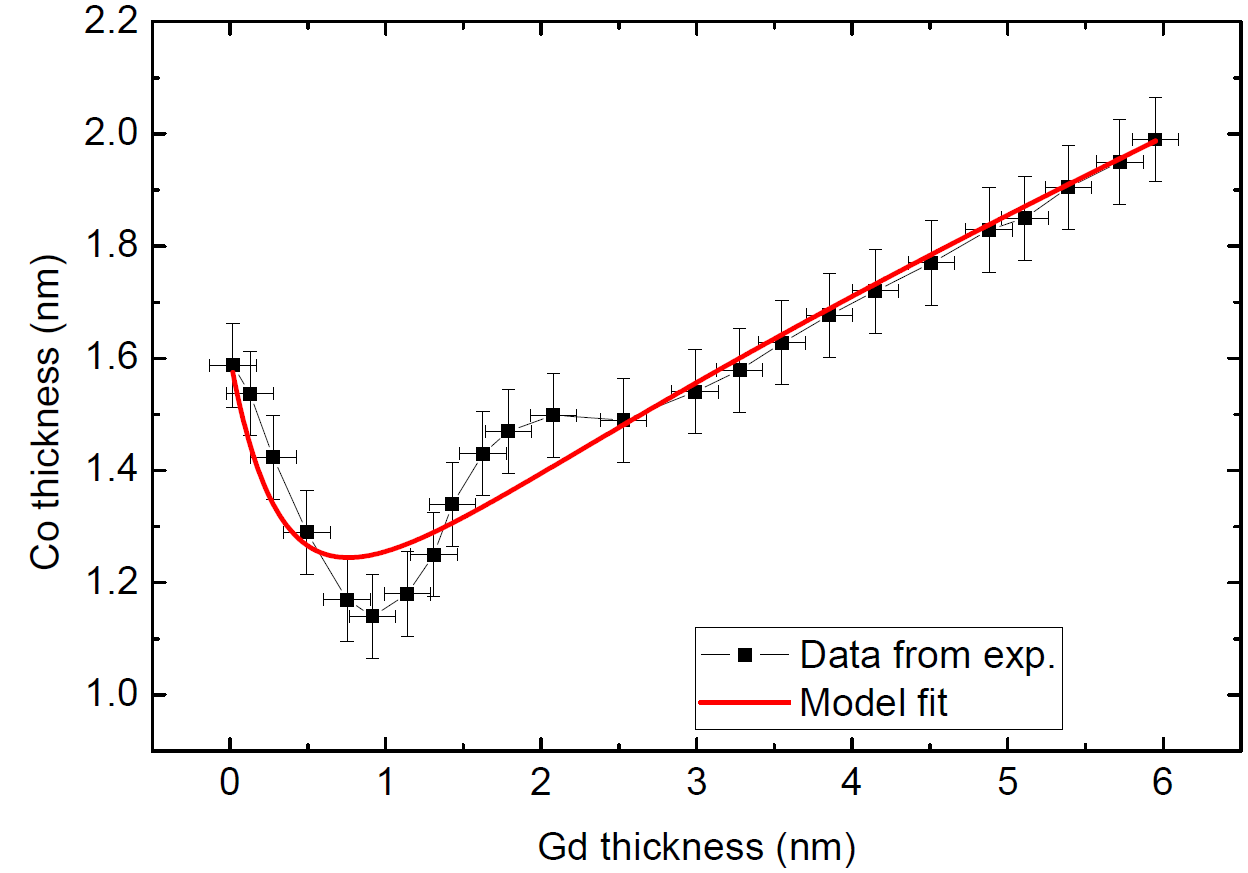}
\caption{\label{fig_S02} Values for the SRT obtained experimentally on a $Ta(4nm)/Pt(4)/$ $Co(t_{Co})/Gd(t_{Gd})/TaN(4)$ sample and used to extract the fitting parameters in Eq. \ref{total_fit}.}
\end{figure}

Here  $x$ and $y$ are, respectively, the Gd and Co thicknesses in the phase diagram of Fig. S\ref{fig_S02}. The value of the parameters used in our model are listed in Table S \ref{tab_material} .

\begin{table}[h!]
    \centering
    \begin{tabular}{||c c c||} 
 \hline
    Parameter & Value  & Description \\ [0.5ex] 
 \hline\hline
 $K_s$ & 1.7 mJ/m$^2$  &	Interfacial anisotropy (from exp.) \\
  \hline
$K_{ME}$ & 0.02 mJ/m$^2$  &	Magnetoelastic anisotropy (from exp.) \\
  \hline

 $M_{Co}$ &  1.3 MA/m &	Cobalt magnetization (from exp.) \\ 
  \hline
  $M_{Gd}$ & 1.4 MA/m & Gadolinium magnetization at Co/Gd interface (from Ref. \cite{kools})  \\ 
 \hline
  $a_{0}$ & 0.13 (-) & Growth parameter of intermixing region (from exp.)  \\ 
 \hline
 $\lambda_K$ & 0.51(15) nm &	Change of PMA energy characteristic decay length (Fit parameter) \\
  \hline

 $\lambda_{Gd}$ &  0.59(22) nm &	Gd magnetization decay characteristic decay length (Fit parameter) \\ 
  \hline
  $\Delta_K$ & 3.96(41)$\times10^{-4}$ J/m$^2$ &	Change of PMA energy (Fit parameter)  \\ 

 \hline
\end{tabular}
\caption{Parameters used in the model for the magnetostatics of uncompensated Co/Gd synthetic ferrimagnets used for the calculations of the SRT. The term $K_{ME}$ is considered zero for when external strain is not applied to the sample. }
\label{tab_material}
\end{table}

The values of $\lambda_K$, $\lambda_{Gd}$ and $\Delta_K$ are instead determined using a numerical fit and are reported in Table S \ref{tab_material}. To fit this equation to the phase diagram obtained experimentally, it is convenient to find the Co-thickness (y) where the anisotropy energy ($E_{tot}$) is equal to zero (spin reorientation transition, SRT). Solving Eq. \ref{total_mix} for y gives:

 \begin{equation} \tag{S.3} \label{total_fit}
 \begin{array}{l}
y_0(x)= \frac{2}{M_{Co}^2 \mu_0}\left(-\left(K_s - \Delta K \left(1-\exp\left(\frac{-x}{\lambda_{K}}\right)\right)\right)\right) - \frac{1}{2}\mu_0 a_0 M_{Co}^2 x \\ -\frac{1}{4}\mu_0 \lambda_{Gd} M_{Gd}^2 \left(1-\exp\left(\frac{-2a_0x}{\lambda_{Gd}}\right)\right) + \frac{1}{4}\mu_0 M_{Gd}^2 \lambda_{Gd} \left(1-\exp\left(\frac{-2x}{\lambda_{Gd}}\right)\right).
\end{array}
\end{equation}

Note that for determining the fit parameters, the measurements were taken without externally applied strain. Accordingly, the magnetoelastic energy term $K_{ME}$ is set to zero in Eq. \ref{total_fit}. The experimental data used for the numerical fit are reported in Fig. S\ref{fig_S02}. The sample used for the numerical fit, explores a wide thickness range (Gd from 0 to 6 nm) in a double wedge fashion to improve accuracy. Consequently, the  dimensions of this sample exceed the 1x1 cm size of the bending device. For this reason, single wedge samples have been deposited for the strain-dependent study. 

The feature around $t_{Gd}$=2 nm in Fig. S\ref{fig_S02} is not captured by out toy model, and might be due to the additional intermixing caused during sputtering, not included in Eq. \ref{total_mix}. 


\subsection*{ \textbf{S2} -Application of strain}

\begin{figure}[h!]
\centering\includegraphics[width=13cm]{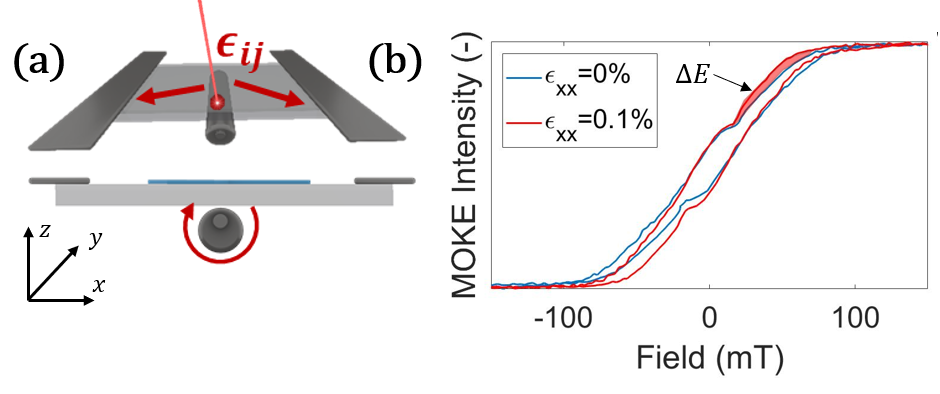}
\caption{\label{fig_S01} (a) schematic of the three point bending method used to externally strain the sample. The strain is mostly uniaxial along the $x$ direction. (b) hysteresis loops measured before (blue) and during (red) application of in-plane strain for a sample of Pt/Co(1.85 nm)/Ta. The area highlighted in red corresponds to the magnetoelastic energy in the strained system. The magnetic field was applied along the OOP direction ($z$).}
\end{figure}

To obtain information about the magnetoelastic properties of the material, the substrate was bent mechanically with a 3 point bending sample holder, as shown schematically in Fig. S\ref{fig_S01} (a). A square sample of 1 by 1 cm is vertically constrained on two sides and pushed uniformly from below by a cylinder that has an off-centered rotation axis. The device generates a tensile strain in the plane of the sample up to $0.1$ $\%$ when the cylinder is rotated by 90$^\circ$. The strain is mostly uniaxial and has been measured with a strain gauge on the substrate surface. 

Magnetic hysteresis loops are recorded before and after the application of the tensile strain and are used to estimate the magnetoelastic anisotropy. As  previously reported\cite{PMA,spinvalve} the magnetic  anisotropy $K_{eff}$ is linked to the energy stored in the magnetization curves. For example the PMA energy is given by the area enclosed between the magnetic loops measured with field along IP and OOP direction.  If then the strain in the film is non-zero, the magneto-elastic coupling contributes in principle to the effective anisotropy. Two hysteresis loops measurements, before and after the application of strain, are sufficient to estimate $K_{ME}$. Indeed the total anisotropy of the system is $K_{eff}=K_{s}$ and $K_{eff}=K_{s}+K_{ME}$ before and after the application of strain, respectively.  The magnetoelastic anisotropy $K_{ME}=-\frac{3}{2}\lambda_s Y \epsilon$ is linked to reversible part of the hysteresis loops (close to the saturation) according to

 \begin{equation} \tag{S.4} \label{eq_strain_eanis}
K_{ME}=M_s \Delta E=-\frac{3}{2}\lambda_s Y \epsilon
\end{equation}

where $\Delta E$ is the anisotropy energy measured by the difference in area below the strained and unstrained curves, $\epsilon$ is the strain $\lambda_s$ is the magnetostriction and Y is the Young's mudulus of the material. In our case $\epsilon=0.1\%$ and Y=200 GPa. $\Delta E$ corresponds to the reversible part, i.e. the red marked area in Fig. S\ref{fig_S01} (b). The value of magnetoelastic anisotropy was calculated using the value of saturation magnetization ($M_s$) of the stack taken from literature and reported in Table S \ref{tab_material}.

\newpage
